\documentclass{aa}

\usepackage{graphicx}
\usepackage{txfonts}

\begin{document}

\title{Quasi-stellar objects and galaxy mass density profiles derived\\using the submillimetre galaxies magnification bias}
\titlerunning{QSOs and galaxy mass density profile with Magnification Bias}
\authorrunning{Crespo D. et al.}

\author{Crespo D.\inst{1,2}, Gonz{\'a}lez-Nuevo J.\inst{1,2}, Bonavera L.\inst{1,2}, Cueli M. M.\inst{1,2}, Casas J. M.\inst{1,2}, Goitia E.\inst{1}}

\institute{$^1$Departamento de Fisica, Universidad de Oviedo, C. Federico Garcia Lorca 18, 33007 Oviedo, Spain\\
$^2$Instituto Universitario de Ciencias y Tecnologías Espaciales de Asturias (ICTEA), C. Independencia 13, 33004 Oviedo, Spain\\
}
\date{Received 13 May 2022; accepted 23 September 2022}
\abstract
{The magnification bias on the submillimetre galaxies (SMGs) is a gravitational lensing effect, where the SMGs are used as background lensed galaxies. This effect can be used to investigate the mass density profiles of different types of foreground lenses.
}
{In this work, we want to exploit the magnification bias of the SMGs using two different foreground samples, quasi-stellar objects (QSOs) and galaxies. Our aim is to study and compare their mass density profiles and estimate their masses and concentrations.
}
{The background SMG sample consists of objects observed by \textit{Herschel} with $1.2<z<4.0$ (mean redshift at $\sim 2.2$). The foreground samples are QSOs with spectroscopic redshifts 0.2 < z < 1.0 (mean redshift at $\sim 0.7$) and massive galaxies with also spectroscopic redshifts 0.2 < z< 1.0 (mean redshift at $\sim 0.3$).
The cross-correlation measurements are estimated with the Davis-Peebles estimator by stacking the SMG--QSO and SMG--galaxy pairs for the two analysed cases, respectively.
The advantage of such an approach is that it allows us to study the mass density profile over a wide range of angular scales, from $\sim 2$ to $\sim 250$ arcsec, including the inner part of the dark-matter halo ($\lesssim 100$ kpc).
Moreover, the analysis is carried out by combining two of the most common theoretical mass density profiles in order to fit the cross-correlation measurements.
}
{The measurements are correctly fitted after splitting the available angular scales into an inner and an outer part using two independent mass density profiles, one for each region. In particular, for the QSOs, we obtain masses of $\log_{10}(M/M_{\odot})=13.51\pm0.04$ and of $\log_{10}(M/M_{\odot})=$13.44$\pm$0.17 for the inner and outer parts, respectively. The estimated masses for the galaxy sample are $\log_{10}(M/M_{\odot})=13.32\pm0.08$ and $\log_{10}(M/M_{\odot})=$12.78$\pm$0.21 for the inner and outer parts, respectively. 
The concentrations for the inner part are much higher than those for the outer region for both samples: $C=6.85\pm0.34$ (inner) and $C=$0.36$\pm$0.18 (outer) for the QSOs and $C=8.23\pm0.77$ (inner) and $C=$1.21$\pm$1.01 (outer) for the galaxies.
}
{
In both samples, the inner part has an excess in the mass density profile with respect to the outer part for both QSOs and galaxy samples. We obtain similar values for the central mass with both samples, and they are also in agreement with those of galaxy clusters results. However, the estimated masses for the outer region and the concentrations of the inner region both vary with lens sample. We believe this to  be related to the probability of galactic interactions and/or the different evolutionary stages.}


\maketitle


\section{Introduction}

Magnification bias, one of the effects of gravitational lensing, is the magnification of the apparent flux densities of background sources that suffer from lensing due to the presence of a foreground lens. Such a boost in the flux might be enough to make the background source detectable in cases where otherwise it would be below the instrument detection limit \citep[e.g.][]{BON21P}. In order to see excess flux density, the  logarithmic slope of the integrated source number counts of the background sources needs to be steep. Estimation of the magnification bias can be performed through the cross-correlation function (CCF), which equates to the excess expected between two source samples with non-overlapping redshift distributions with respect to the case with no magnification \citep{Scr05,Men10,Hil13,Bar01}. This effect was directly measured by \citet{DUN20} with the Atacama Large Millimetre Array (ALMA).
\newline
The submillimetre galaxies (SMGs) are considered \citep{GON14,GON17} to be an optimal sample of background sources for magnification bias studies thanks to their steep source number counts, $n(>S) = n_0\,S^{-\beta}$ with $\beta \gtrsim 2$, and their high redshift, $z>1$. Moreover, the foreground lens (detected in the optical band) emission at submillimetre wavelengths can be considered negligible, while the SMGs are invisible in the optical band for a similar reason. Therefore, both samples are not cross-contaminated by each other.
For these reasons, SMGs have already been successfully used in magnification bias analyses to investigate the projected mass density profile and concentration of foreground samples of quasi-stellar objects \citep[QSOs;][]{BON19}, for cosmological studies \citep{BON20, GON21, BON21}, and to observationally constrain the halo mass function \citep{CUE21, CUE22}.

With respect to the type of lens discussed in this work,  QSOs are extremely luminous active galactic nuclei (AGN) that can be detected over a very broad range of distances. They are very suitable as background objects in lensing studies in general and also when adopting CCF measurements for magnification bias studies \citep[e.g.][]{Bar93,Scr05,Men10}.
For example, the Sloan Digital Sky Survey (SDSS) Quasar Lens Search (SQLS) identified 28 galaxy-scale multiply imaged quasars that undergo gravitational lensing \citep{Ogu06,Ogu08}.
On the other hand, the first strong gravitational lensing caused by a QSO (SDSS J0013+1523 at z = 0.120) was detected by \cite{COU10}. Additional cases of QSOs acting as foreground objects have since been identified \citep[e.g. ][]{COU12, Har15, Dan17}.
QSOs are also used in shear measurements as in \citet{Man09} and \citet{Luo22}, where the authors measure the weak-lensing shear distortions taking AGNs as lenses, making these studies complementary to magnification bias measurements.
QSOs acting as lenses on SMGs have also been used to extract information on the mass density profile \citep[see][]{BON19} with the cross-correlation approach.

Large galaxy surveys \citep[as SDSS, ][]{YOR00Y} are undoubtedly very useful in cosmology, \citep[e.g. BOSS][]{DAW13}. A first attempt at measuring lensing-induced cross-correlations between SMGs and low-z galaxies was carried out by \cite{WAN11}, who found convincing evidence of the effect. With much better statistics, this bias was studied in detail with the CCF between SMGs and massive galaxies by \cite{GON14, GON17}. More recently, the magnification bias on SMGs  produced by such massive optical galaxies at $z<<1$ was analysed to derive complementary and independent constraints on the main cosmological parameters \citep{BON20, GON21, BON21}.

Galaxy clusters, instead, are massive bound systems that can be used to track the large-scale structure of the Universe because they are generally placed in the knots of filamentary structures. They are also used for cosmological studies \citep[e.g.][]{ALL11}  to investigate the evolution of galaxies \citep[][]{DRE80,BUT78,BUT84,GOT03} and the lensed high-redshift galaxies \citep[e.g.][]{BLA99}. Moreover, the correlation between galaxy clusters  and selected background objects has been exploited to investigate possible lensing effects \citep[][]{MYE05,LOP08} and, in particular, the CCF measurements using the SMGs as background sample to estimate the masses and concentration of the galaxy clusters \citep[][]{FER22}.

Although the signal produced by weak lensing events is fainter than that produced by strong lensing events, the former are more frequent, allowing them to be studied by applying stacking techniques.
Such techniques consist in co-adding the signal of weak or undetectable objects in order to enhance the signal and lower the background emission so that the overall statistical properties of this signal can be obtained. Stacking has been used in the recovery of the integrated signal of the Sachs–Wolfe effect with \textit{Planck} data \citep{Pla14,Pla16b}, and in studies of the faint polarised signal of radio and infrared sources \citep[see][]{Stil14, BON17a, BON17b}.
Moreover, the stacking technique has been used to obtain the mean spectral energy distribution (SED) of optically selected quasars \citep{Bia19} and to recover the weak gravitational lensing of the cosmic microwave background in the \textit{Planck} lensing convergence map \citep{Bia18}.
Stacking has also been used to probe star formation in dense environments of $z\sim 1$ lensing halos \citep{Wel16} and more recently in \cite{FER22} to recover the CCF signal between galaxy clusters for different richness ranges and SMGs.

In this work, the stacking technique is exploited to estimate and compare the mass density profiles of three different types of lenses (QSOs, galaxies, and galaxy clusters) and is organised as follows. Sections \ref{sec:data} and \ref{sec:method} present the data used and the methodology we are applying (both the stacking and the CCF estimation), respectively. Section \ref{sec:framework} describes the theoretical framework of the CCF, weak gravitational lensing, and halo mass density profiles. The results are discussed in Sect. \ref{sec:results} and the conclusions are summarised in Sect. \ref{sec:concl}. The cosmological model used throughout the paper is the flat $\Lambda$ cold dark matter ($\Lambda$CDM) with the cosmological parameters estimated by \citet[][$\Omega_m = 0.31$, $\sigma_8 = 0.81$ and $h = H_0 /100$ km s$^{-1}$ Mpc$^{-1} = 0.67$]{PLA18_VI}.

 \begin{figure}[ht]
 \includegraphics[width=0.49\textwidth]{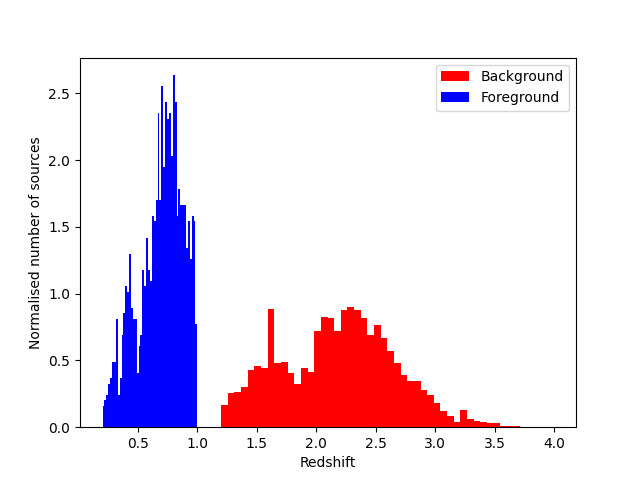} 
 \caption{Redshift distributions of the background sources from the H-ATLAS sample (red histogram), specifically the DR1 and NGP zones, and the QSO sample acting as lenses from the DR7 \citep{SCH10} and DR12 \citep{PAR17} catalogues (blue histogram).}
 \label{fig:qso_histograms}
 \end{figure}
 
\section{Data}
\label{sec:data}
\subsection{Foreground samples}
In this work, we estimate the CCF between two different lens samples (QSOs and galaxies) and the SMGs. Our initial sample of QSOs was obtained from that used in \cite{BON19}. This QSO sample is selected from the Sloan Digital Sky Survey (SDSS), in particular, SDSS-II and SDSS-III Baryon Oscillation Spectroscopic Survey (BOSS) catalogues of spectroscopically confirmed QSOs detected over 9376 deg$^2$. We use the 7th\footnote{Available at \tiny{\url{http://classic.sdss.org/dr7/products/value_added/qsocat_dr7.html}}.} \citep[DR7,][]{SCH10} and 12th\footnote{Available at \tiny{\url{http://www.sdss.org/dr12/algorithms/boss-dr12-quasar-catalog/}}.} \citep[DR12,][]{PAR17} SDSS data releases \citep[see][for a detailed discussion of the QSO target-selection process]{ROS12}. The selection is based mainly on the DR7 catalogue, which mostly includes `low-$z'$ sources at $z<2.5$. The DR12 sample specifically targeted QSOs at $z>2.5$. Even so, there is a secondary maximum around $z\sim 0.8$ as a consequence of colour degeneracy in the target selection of photometric data that leads to the observation of a fraction of low-redshift QSOs. Approximately 4\% of the DR7 objects at $z > 2.15$ were re-observed for DR12, and so a combined sample was created joining the QSO information contained in the DR12 sample, namely any QSO present in both DR7 and DR12 is included only once.

In order to minimise the potential cross-contamination due to the redshift overlap between the foreground and background samples, we selected only QSOs with redshift between $z=0.2$ and $z=1.0$. This restriction leaves a total of 1546 QSOs in the common area with the background sample. The redshift distribution of the selected QSOs is shown in Fig. \ref{fig:qso_histograms} through the blue histogram with a mean redshift $\left\langle z\right\rangle =0.7^{+0.1}_{-0.2}$ (the uncertainty indicates the $1\sigma$ limits).

On the other hand, the foreground galaxies sample was extracted from the GAMA II \citep{DRI11,BAL10,BAL14} survey, which was coordinated with the H-ATLAS one \citep[our background sample; see following subsection;][]{PIL10,EAL10} to maximise the common area. Indeed, they both covered the three equatorial regions at 9, 12, and 14.5 h and part of the south Galactic pole, amounting to a common area of $\sim207\text{deg}^2$.

\begin{figure}[ht]
 \includegraphics[width=0.49\textwidth]{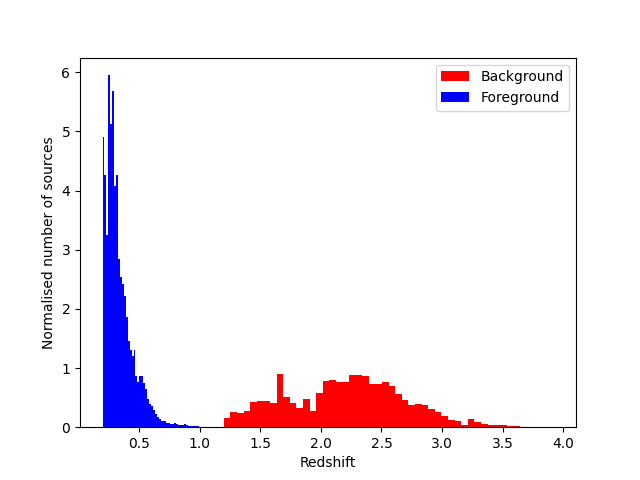} 
 \caption{Redshift distributions of the background sources from the H-ATLAS sample (red histogram), only the DR1 zone, and the galaxies sample acting as lenses from the GAMA catalogue \citep[][blue histogram]{LIS15}.}
 \label{fig:glx_histograms}
 \end{figure}

The foreground galaxy sample consists of GAMA II sources within the interval $0.2<z<0.8$, which results in approximately 102672 galaxies (significantly higher than the case of QSO) with a mean spectroscopic redshift of $\left\langle z\right\rangle =0.3^{+0.1}_{-0.1}$. Their redshift distribution is shown in Fig. \ref{fig:glx_histograms} in blue. The red histogram corresponds to the background sample and is not exactly the same as in Fig. \ref{fig:qso_histograms} due to the different zones covered by both foreground samples.

\subsection{Background sample}
The background sample used in this work is the same for all the different foreground samples, and consist of the officially detected galaxies from \textit{Herschel Astrophysical Terahertz Large Area Survey} (H-ATLAS) data. These observations were collected by the \textit{Herschel Space Observatory} \citep{PIL10}, cover about 610 deg$^{2}$, and are composed of five different fields. Three of these fields are taken from the first data release on the celestial equator, known as Galaxy and Mass Assembly (GAMA) fields or Data Delivery 1 \citep[DR1,][]{VAL16,BOU16,RIG11,PAS11,IBA10}, and correspond to equatorial regions at 9, 12, and 14.5 h. 

The other fields correspond to the North and South Galactic Poles, and are referred to as Data Delivery 2 (DR2), or NGP and SGP, respectively \citep{SMI17,MAD18}. These latter two have total areas of 180.1 deg$^2$ and 317.6 deg$^2$, respectively. We use the three H-ATLAS GAMA fields (G09, G12 and G15) for both foreground samples. In the case of the QSO sample, only the NGP field is covered and only the SGP field is covered in the case of the galaxy sample. However, we decided to discard the SGP measurements due to the large uncertainty, which is much larger than that estimated in the GAMA fields. The main reason is that the galaxy distribution is very inhomogeneous within the field. Although taking into account this effect in the random simulations following \citet[][]{GON21} improved the results, we considered the results from this field to be model dependent and not sufficiently robust to be considered for further analysis.

The H-ATLAS is composed of two instruments, the Photodetector Array Camera and Spectrometer (PACS) \citep[PACS;][]{POG10} and the Spectral and Photometric Imaging REceiver \citep[SPIRE;][]{GRI10}, which work in five photometric bands: 100, 160, 250, 350, and 500 $\mu$m. In both H-ATLAS DRs, there is an implicit 4$\sigma$ detection limit at 250 $\mu$m ($\sim S_{250} \gtrsim 29$ mJy). The 1$\sigma$ noise for source detection, including both confusion and instrumental noise, is 7.4mJy at 250 $\mu$m \citep{VAL16,MAD18}. In addition, as in \citet{GON17}, a 3$\sigma$ limit at 350 $\mu$m was applied to increase the robustness of the photometric redshift estimation.
 
To avoid any overlap in the redshift distribution between the lens and background samples, only those SMGs with photometric redshift between 1.2 and 4.0 are taken into account. Such redshifts were estimated by taking the mean of a minimum $\chi^2$ fit of a template SED to SPIRE data and also using PACS data when possible. The SED of SMM J2135-0102
 \citep[‘The Cosmic Eyelash’ at z = 2.3;][]{IVI10,SWI10} was shown to be a good template; it was found to be the best overall template with $\Delta z/(1 + z) = -0.07 $ and a dispersion of $0.153$ \citep{IVI16,GON12,LAP11}. Following this selection process, we have 49293 and 27517 SMGs for the QSOs and galaxy sample analysis, respectively. They constitute approximately 21 and 23 per cent of the corresponding initial background samples (the sources in the common areas between the SMGs sample and the QSOs and galaxy samples, respectively).
 
 The redshift distribution of the background samples (in red) is shown in Fig. \ref{fig:qso_histograms} for the QSO sample and Fig. \ref{fig:glx_histograms} for the galaxy sample. Although the associated redshift distribution is not exactly the same  for both cases, they have the same mean redshift $\left\langle z\right\rangle =2.2^{+0.4}_{-0.5}$ (the uncertainty indicates the $1\sigma$ limits).

\section{Measurements}
\label{sec:method}

\subsection{Stacking}
\label{sec:stack}

The technique known as ‘stacking’ is a method that can be used to determine the mean flux density of a large set of sources that are too weak to be analysed individually \citep[][]{Dol06,Mar09,Bet12}. This tool is of value in our particular science case because it provides us with valid statistical information in cases where the noise is comparable to the signal, preventing detection. It consists in adding up patches of interest of the sky in order to enhance the signal, which would be otherwise undetectable in single events, not only for  statistical flux density measurements.

A slight variation of this method was previously used by \cite{BON19} and \cite{FER22} to study the CCF signal due to magnification bias between SMGs (as the background sample) and QSOs and galaxy clusters (as the foreground sample), respectively. In these cases, the authors stacked the position of the sources around the lenses, not their flux densities, because the signal of interest is the number of background sources near the lens positions. \cite{BON19} derived the stacked magnification bias of lensed SMGs in lens positions signposted by QSOs and \cite{FER22} study the stacked magnification bias produced by galaxy clusters acting as lenses on the background SMGs. 

The same procedure is applied in this work and follows a similar approach to the traditional CCF estimator, but with the advantage that it accounts for positional errors and can be used to identify the foreground--background pairs in the stacked map. Here we use this procedure to search for background sources in a circular area centred on the lens position. The search radius is set to be 200 arcsec, which means we must build a square map of 0.5 arcsec pixel size and 400 $\times$ 400 pixels centred at the lens position (QSO or galaxy, depending on the case being analysed) containing the paired background sources to the lens. The maps obtained for all the lenses are added up and normalised to the total number of lenses (102672 targets for the galaxy sample and 1546 for the QSO sample) to produce the final stacked map. To account for the positional accuracy of the catalogues, a Gaussian filter of $\sigma=2.4$ arcsec is applied. This value corresponds to the positional accuracy of the H-ATLAS catalogue \citep[SMGs;][]{BOU16,MAD18} and is more than an order of magnitude greater than that for the QSO or galaxy samples.

\begin{figure*}[ht]
\includegraphics[width=0.49\textwidth]{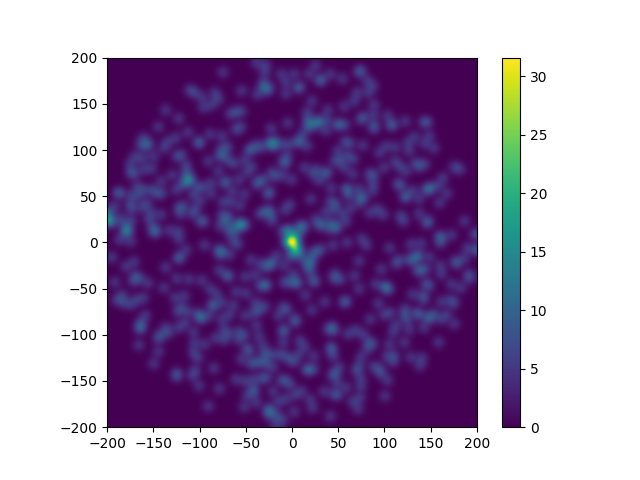}
\includegraphics[width=0.49\textwidth]{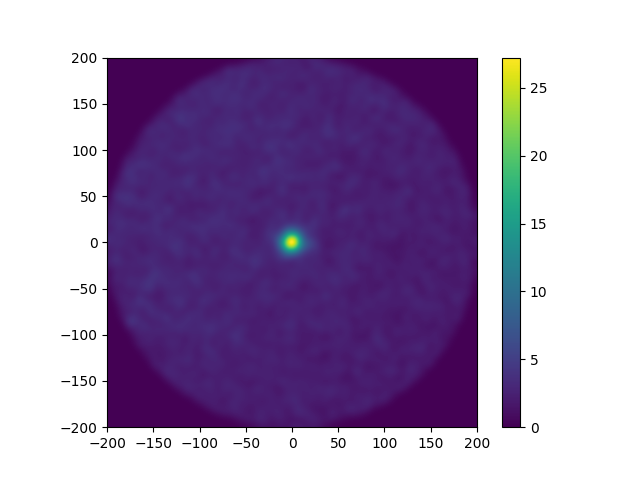}
 \caption{Relative excess probability (stacked image/random mean-1) for QSO (left) and galaxy (right) pairs considering the background sources within an angular radius of 200 arcsec from the position of the target. In both cases, the pixel size is 0.5 arcsec and we apply a $2.4\sigma$ Gaussian filter to take into account the positional uncertainties (see text for more details). The mean and standard deviation per pixel of the random stacked image for QSOs are $2.25 \times 10^{-6}$ and $1.21\times 10^{-6}$, respectively, whereas the mean and standard deviation per pixel of the random stacked image for the galaxy sample are $2.32 \times 10^{-6}$ and $1.15\times 10^{-6}${, respectively}.
 }
 \label{fig:stack_map}
\end{figure*}

\begin{figure*}[ht]
    \includegraphics[width=0.49\textwidth]{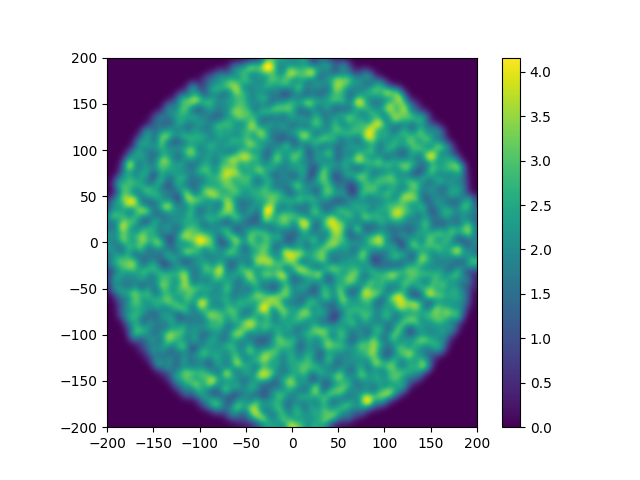}
    \includegraphics[width=0.49\textwidth]{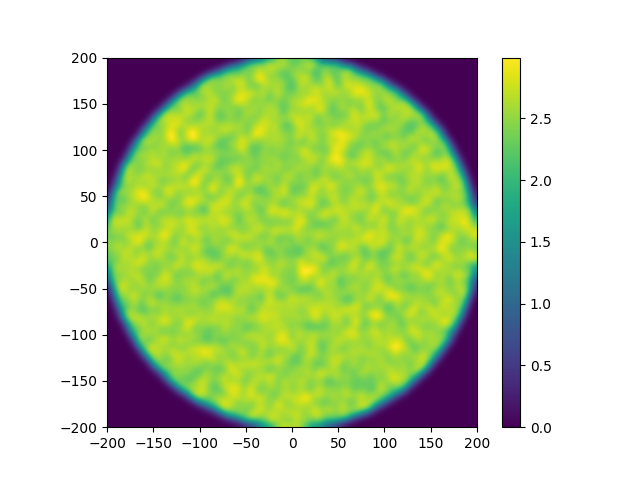}
 \caption{Relative excess probability (stacked image/mean-1) of random pairs for QSOs (left) and galaxies (right) using the same radius and pixel size as for the data case and smoothed with the same Gaussian filter.
 }
 \label{fig:random_map}
\end{figure*}

The resulting maps for QSO and galaxy samples are shown in the left and right panels of Fig. \ref{fig:stack_map}, respectively. The colour scale indicates the relative excess probability with respect to the random mean value (stacked pairs/random mean-1) of finding a background--foreground pair. The mean and standard deviation per pixel of the random stacked image are $2.25\times10^{-6}$ and $1.21\times10^{-6}$ for QSOs and $2.32\times10^{-6}$ and $1.15\times10^{-6}$ for the galaxies.
Figure \ref{fig:random_map} shows the expected signal in the absence of lensing where QSOs and galaxy positions are randomly simulated (left and right panel, relatively) following the same procedure as for real data. The random stacked maps were simulated using approximately 6\,000 and 200\,000 targets for the QSOs and galaxy samples for each GAMA region and 14\,000 targets for the NGP region for QSOs (ten times the round-up value of the available possible foreground lenses).

\subsection{Cross-correlation function}
\label{sec:xcorr}

We applied the CCF estimator as in \cite{BON19} to the stacked images in order to extract statistical information about the lens sample and the lensing system in general. The adopted estimator is that provided by \cite{Dav83}:
\begin{equation}
    \tilde{w}_{x}(\theta)=\frac{\text{DD}}{\text{RR}}-1
    \label{eq:wx}
,\end{equation}
where the measurements are computed by drawing a set of circles the centres of which are at the central position of the map and the radii of which logarithmically increase by 0.05 starting from 1 arcsec. The values of the pixels contained in the initial circle and in each one of the rings are added up as $\text{DD}$ and the same procedure is repeated for the random map to produce $\text{RR}$.

To compute the CCF errors, each one of the first four rings is divided into sections of equal area with no more than eight pixels per section.
The following rings instead have enough pixels to be split in 15 sections of equal area. Following \cite{BON19}, in each ring, a Jackknife method is applied to estimate the $\text{DD}$ and $\text{RR}$ uncertainties, which are then propagated according to Eq. \eqref{eq:wx} to get the errors on $\tilde{w}_{x}$. Moreover, we analyse the possible correlation between the data points by computing the covariance matrix when performing the Jackkinfe analysis; this results in a negligible correlation.

Figures \ref{fig:qso_xcorr} and \ref{fig:glx_xcorr} show the estimated CCFs using stacking (black dots) and the CCF estimated following \citet[][red dots]{GON17} for the QSOs and galaxy samples, respectively. The angular scales range from 1 to 1000 arcsec, allowing us to study the mass density profile over a wide spectrum of physical scales ($\sim 5$ kpc to $\sim 4$ Mpc for $z=0.3$ or $\sim 8$ kpc to $\sim 7$ Mpc for $z=0.7$): from the central region (mainly related with strong lensing events) to the outer part of the lens object (where weak lensing is most likely to happen). This is shown in Fig. \ref{fig:stack_map} and measured in Figs. \ref{fig:qso_xcorr} and \ref{fig:glx_xcorr}, where the CCF signal is stronger at the smaller angular scales and decreases towards the larger ones.

\section{Theoretical framework}
\label{sec:framework}

The cross-correlation measurements described above are strongly related to the physical properties of the mass distribution of the lenses. Indeed, the magnification field due to the density profile of a lens can be computed and linked directly to the cross-correlation signal by interpreting the physical meaning of the latter.

\subsection{Gravitational lensing and the cross-correlation function}
\label{sec:theory}

One of the consequences of the phenomenon of gravitational lensing is the modification of the integrated number counts of a sample of background sources due to a mass distribution between this latter sample and the observer. Indeed, due to the competing effects of a magnification that enhances fainter sources and a dilution that enlarges the solid angle in the sky, the number of background sources per solid angle and redshift  with observed flux density larger than $S$ is modified at every two-dimensional angular position $\vec{\theta}$ in the sky as \citep{Bar01}
\begin{equation*}
    n(>S,z;\vec{\theta})=\frac{1}{\mu(\vec{\theta})}n_0\Big(>\frac{S}{\mu(\vec{\theta})},z \Big),\label{numbercounts1}
\end{equation*}
where $\mu(\theta)$ is the magnification field and $n_0$ denotes the integrated number counts in the absence of lensing. If the (unlensed) background source number counts are assumed to follow a redshift-independent power-law behaviour, that is, $n_0(>S,z)=A\,S^{-\beta}$, then it is clear that
\begin{equation*}
\frac{n(>S,z;\vec{\theta})}{n_0(>S,z)}=\mu^{\beta-1}(\vec{\theta}).
\end{equation*}
The connection between the magnification field of a sample of lenses and the cross-correlation observable that we aim to measure can be elucidated when we understand the physical meaning of the above ratio. Indeed, the quantity $n(>S,z_b;\vec{\theta})/n_0(>S,z_b)$ represents the excess (or lack) of background sources (in direction $\vec{\theta}$ as viewed by a lens at $z_l$) at redshift $z_b>z_l$ with respect to the case without lensing. The angular CCF between a sample of foreground lenses at redshift $z_l$ and a sample of background objects at redshift $z_b$ is defined as
\begin{equation*}
    w_x(\vec{\theta};z_l,z_b)\equiv\langle\delta n_f(\vec{\phi})\,\delta n_b(\vec{\phi}+\vec{\theta})\rangle,
\end{equation*}
where $\delta n_b$ and $\delta n_f$ denote the background and foreground object density contrast, respectively. It follows from the above argument that
\begin{equation*}
    w_x(\vec{\theta};z_l,z_b)=\mu^{\beta-1}(\vec{\theta})-1,
\end{equation*}
as we are stacking the lenses at a fixed position, which we take as the origin. We thus obtained a relation between our observable and a quantity that depends on the physical properties of the mean density profile of the lenses.

\subsection{Mass density profiles}
\label{sec:profiles}

If a lens located at an angular diameter distance $D_d$ (from the observer) deflects the light coming from a source at an angular diameter distance $D_s$, the convergence field at an angular position $\vec{\theta}=\vec{xi}/D_d$ on the image plane is defined as
\begin{equation*}
    \kappa(\vec{\theta})=\frac{\Sigma(D_d\vec{\theta})}{\Sigma_{\text{cr}}},
\end{equation*}
where $\Sigma(\vec{\xi})$ denotes the mass density projected onto a plane perpendicular to the light ray and
\begin{equation*}
    \Sigma_{\text{cr}}=\frac{c^2}{4\pi G}\frac{D_s}{D_dD_{ds}}
\end{equation*}
is the critical mass density, where $D_{ds}$ is the angular diameter distance between the lens and the background source.

In the case of axially symmetric lenses, choosing the origin as the symmetry centre yields $\kappa(\vec{\theta})=\kappa(\theta)$ and the magnification field $\mu(\theta)$ is related to the convergence via \citep{Bar01}
\begin{equation*}
    \mu(\theta)=\frac{1}{(1-\bar{\kappa}(\theta)(1+\bar{\kappa}(\theta)-2\kappa(\theta))},
\end{equation*}
where $\bar{\kappa}(\theta)$ is the mean surface mass density within the angular radius $\theta$. We now present the expressions of the magnification field for the two models we consider for the mean density profile of the lenses.

\subsubsection{Navarro-Frenk-White profile}

The Navarro-Frenk-White \citep[NFW,][]{NAV96} profile is a two-parameter model given by
\begin{equation*}
    \rho_{\text{NFW}}(r;r_s,\rho_s)= \frac{\rho_s}{(r/r_s)(1+r/r_s)^2},
\end{equation*}
where $r_s$ and $\rho_s$ are the so-called scale radius and density, respectively. If halos are identified as spherical overdense regions with a mean density value of $\rho_h$, then $r_s$ and $\rho_s$ are related via
\begin{equation*}
    \rho_s=\frac{\rho_h}{3}\frac{C^3}{\ln{(1+C)}-C/(1+C)},
\end{equation*}
where $C\equiv r_h/r_s$ is referred to as the concentration parameter and $r_h$ is the (truncation) radius of the identified halo.

It can be shown that the NFW profile satisfies \citep{SCH06}

\begin{equation*}
    \kappa_{\text{NFW}}(\theta)=\frac{2r_s\rho_s}{\Sigma_{\text{cr}}}f(\theta/\theta_s)\quad\quad\quad \bar{\kappa}_{NFW}(\theta)=\frac{2r_s\rho_s}{\Sigma_{cr}}h(\theta/\theta_s),
\end{equation*}
where $\theta_s\equiv r_s/D_d$ is the angular scale radius,
\begin{equation*}
    f(x)\equiv \begin{cases} \frac{1}{x^2-1}-\frac{\arccos{(1/x)}}{(x^2-1)^{3/2}}\quad\quad&\text{if } x>1\\
    \,\,\frac{1}{3} \quad\quad&\text{if } x=1\\
    \frac{1}{x^2-1}+\frac{\text{arccosh}(1/x)}{(1-x^2)^{3/2}} \quad\quad&\text{if } x<1
    \end{cases}
\end{equation*}
and
\begin{equation*}
    h(x)\equiv \begin{cases} \frac{2}{x^2}\Big(\frac{\arccos{(1/x)}}{(x^2-1)^{1/2}}+\log{\frac{x}{2}}\Big)\quad\quad&\text{if } x>1\\
    \,{\scriptstyle 2\,(1-\log{2})} \quad\quad&\text{if } x=1\\
    \frac{2}{x^2}\Big(\frac{\text{arccosh }(1/x)}{(1-x^2)^{1/2}}+\log{\frac{x}{2}}\Big) \quad\quad&\text{if } x<1
    \end{cases}.
\end{equation*}

\subsubsection{Singular isothermal sphere profile}
The singular isothermal sphere (SIS) model is given by
\begin{equation*}
    \rho_{\text{SIS}}(r)=\frac{\sigma_v^2}{2\pi G r^2},
\end{equation*}
corresponding to a system of particles with a Maxwell velocity distribution at every radius with one-dimensional velocity dispersion $\sigma_v$, given by
\begin{equation*}
    \sigma_v=\sqrt{\frac{GM}{2r_h}},
\end{equation*}
where $M$ is the mass of the halo described by the profile. The convergence and mean surface density inside $\theta$ are easily shown to be \citep{SCH06}
\begin{equation}
    \kappa_{\text{SIS}}=\frac{\theta_E}{2|\theta|}\quad\quad\bar{\kappa}_{\text{SIS}}(\theta)=\frac{\theta_E}{|\theta|},
\end{equation}
where
\begin{equation}
    \theta_E=4\pi\,\bigg(\frac{\sigma_v}{c}\bigg)^2\frac{D_{ds}}{D_s}
\end{equation}
is the so-called Einstein radius of the model.

\section{Results}
\label{sec:results}

As in \cite{FER22}, with a view to analyse the measured CCFs and to extract physical conclusions, the data have been fitted by a combination of the two mass density profiles 
introduced above. The different fits to the data produced in this work clearly show that a single mass density profile is unable to fit the data at all scales. On the whole, using a combination of two common mass density profiles gives us a better description of the results.

In particular, in the following sections we present our results with CCF measurements using QSOs (Sect. \ref{sec:QSO}) and galaxies (Sect. \ref{sec:GLX}) as lenses. The background objects are from the same SMG sample for both cases.

\subsection{Quasi-stellar objects}
\label{sec:QSO}

Although \cite{BON19} have already performed an analysis of CCF results using stacking with QSOs as lenses, we are driven to repeat this analysis in light of the fact that \cite{BON19} and \cite{FER22} use slightly different theoretical frameworks. 
The \cite{BON19} analysis with QSOs relies on a weak-lensing approximation which is no longer valid at small angular separations (the strong lensing regime). \cite{FER22}, on the other hand, attempted to include the strong lensing regime in their study of galaxies clusters and the weak-lensing approximation is removed from their theoretical framework.
In the present work, we are applying this latter approach ---which does not require any kind of approximation and is therefore valid for all the lensing regimes--- in order to study and compare the cases of galaxy clusters, galaxies, and QSOs acting as lenses, including those in the small angular scales regime. Therefore  a re-analysis of QSOs is needed in order to apply a consistent methodology for all three cases.

\begin{figure*}[ht]
 \centering
 \includegraphics[width=0.49\textwidth]{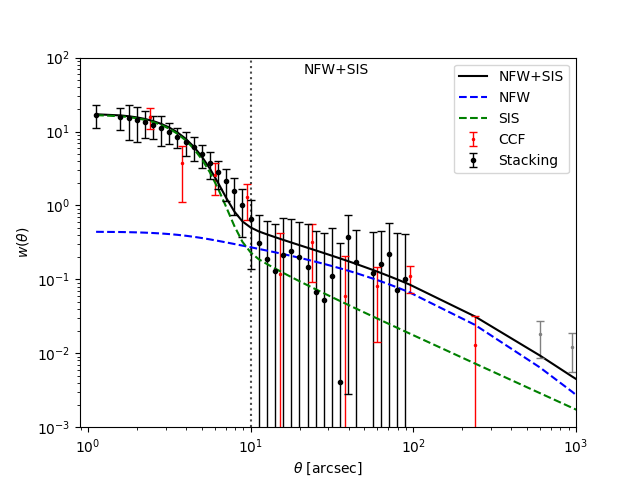}
 \includegraphics[width=0.49\textwidth]{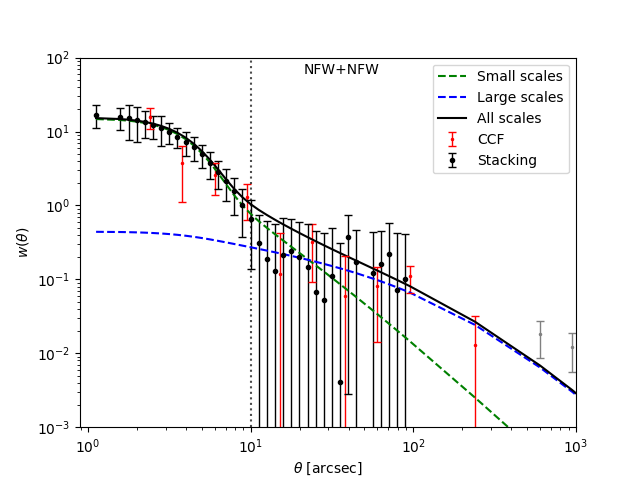}
 \caption{Stacking data (black points) and those of the cross-correlation function (red points) computed using the QSOs acting as lenses as in \cite{BON19}. On the left, the black line represents the NFW+SIS fits, where the corresponding SIS (green dashed line) and NFW (blue dashed line) fits are also separately shown. In the same way, on the right, the NFW+NFW profile is represented by the black line. In this case, the green dashed line corresponds to the fit to the points at the small scales only and the blue dashed line is for large scales. The grey dashed vertical line at $\sim 10$ arcsec represents the separation between small and large angular scale regimes. Grey points are considered outliers and are not taken into account for the analysis.}
 \label{fig:qso_xcorr}
\end{figure*}

The results obtained with stacking are shown in Fig. \ref{fig:qso_xcorr} with black dots. The red points are the CCF measurements derived with the traditional estimator described in \cite{GON17} (which does not make use of the stacking technique). Overall, there is good agreement between the two CCF estimations, with larger uncertainties at larger scales for the stacking approach. The fact that the CCF with no stacking is more robust in the low-signal regime was pointed out by \cite{BON19}. 
The dotted vertical line indicates the two angular separation regions that divide the analysis: the inner part that extends until $10$ arcsec and the outer part for angular scales greater than $10$ arcsec. This particular angular separation was chosen as representative of the transition between the weak and strong lensing regimes. The stacking estimations have larger error bars in the outer region than in the inner one. In particular, this means that the stacking measurements (black points) are less informative above $100$ arcsec than the CCF measurements (red points). For this reason, at angular scales above $100$ arcsec we rely only on the red points to perform our analysis.
Moreover, we exclude the CCF points at the largest scales  from the analysis (grey points in Figs. \ref{fig:qso_xcorr} and \ref{fig:glx_xcorr}, $\theta \gtrsim 200$ arcsec or $\sim 1.4 Mpc$ at $z\sim0.7$ for the QSOs and $\theta \gtrsim 300$ arcsec or $\sim 1.3 Mpc$ at $z\sim0.3$ for the galaxies), because they correspond to the two-halo term. In the case of QSOs, given the shortage of points above 100 arcsec we opt to keep the point at $\theta \sim 250$ arcsec in order to perform the fit, even if it is placed at a scale where the two-halo term might already come into place. Therefore, the estimated concentration value can be considered a lower value of the real concentration. Larger samples, which imply smaller uncertainties, and a proper two-halo term contribution estimation will be needed to improve the accuracy on the concentration at larger scales.

\begin{table}[ht]
\begin{tabular}{ccccc}

\hline \hline
 & \small{$\log_{10}(M/M_{\odot})$}  & \small{$\Delta\log_{10}(M/M_{\odot})$} & \small{$C$}          & \small{$\Delta C$}   \\ 
                  \hline
NFW$_{small}$     & $13.51$       & $0.04$  & $6.85$ & $0.34$ \\
SIS$_{small}$     & $12.934$ & $0.001$  & ---         & ---         \\
NFW$_{large-CCF}$ & 13.44       & 0.17  & 0.37 & 0.18 \\ \hline \hline

\end{tabular}
\caption{\label{table_qso} Derived parameter values for the QSOs for each profile (first column) and angular separation regime (below and above $\sim 10$ arcsec). The second and third columns are the mass and its error and the fourth and fifth are the concentration and its error (only for the NFW profile).}
\end{table}

The left panel of Fig. \ref{fig:qso_xcorr} shows the combined SIS and NFW
profiles with the black solid line. The corresponding individual
NFW and SIS profiles are also shown with dashed red and gray lines. In the overall fit is good, with the SIS part fitting the inner region and the NFW fitting the outer region. This behaviour was not imposed by the methodology, but comes naturally. The major disagreement is in the region between 6 and 10 arcsec where the fit is below the data, as in \cite{FER22}.

Table \ref{table_qso} summarises our results: from left to right, the columns are the profile, the characteristic halo mass and its error, and the concentration $C$ and its error.
In the inner region we obtain a mass value of  $\log_{10}(M/M_{\odot})=12.934\pm0.001$ for the SIS profile. For the outer part (using a NFW profile) the results are $\log_{10}(M/M_{\odot})=$13.44$\pm$0.17 and $C=$0.36$\pm$0.1804, with a negligible correlation between the two parameters (3\%).

It the right panel of Fig. \ref{fig:qso_xcorr}, the fit is obtained with two independent NFW profiles (black solid line). In this case, the fit perfectly explains the measurements, although there is a hint of a small overfit of the data in the region between 10 and 30 arcsec even if still compatible with the large error bars. The NFW fit in the inner region gives a mass value of $\log_{10}(M/M_{\odot})=13.51\pm0.04$ and a concentration of $C=6.85\pm0.34$, with a negligible correlation between the two parameters (1\%). The outer fit using a NFW profile is the same as in the previous case.  

As a comparison, \cite{BON19} estimated a total mass of $M=1.7^{+2.1}_{-0.5}\times 10^{14}M_\odot$ using the NFW profile that is three to four times our current estimates. This difference confirms the importance of improving the theoretical background in order for the analysis to function without any approximation, and demonstrates the effectiveness of the methodology, namely using two different independent mass density profiles for the inner and outer regions.

\subsection{Galaxies}
\label{sec:GLX}

We repeated the same analysis for the sample of galaxies acting as lenses. However, this time the separation between the inner and outer angular separation regions is set at $20$ arcsec (indicated as a dotted vertical line), because the mean redshift of this sample is different from that of QSOs ($<z>\sim0.7$ for QSOs and $\sim 0.3$ for the galaxies) implying different angular separations for roughly the same physical transition scale, $\sim 80$ kpc.

\begin{figure*}[ht]
 \centering
 \includegraphics[width=0.49\textwidth]{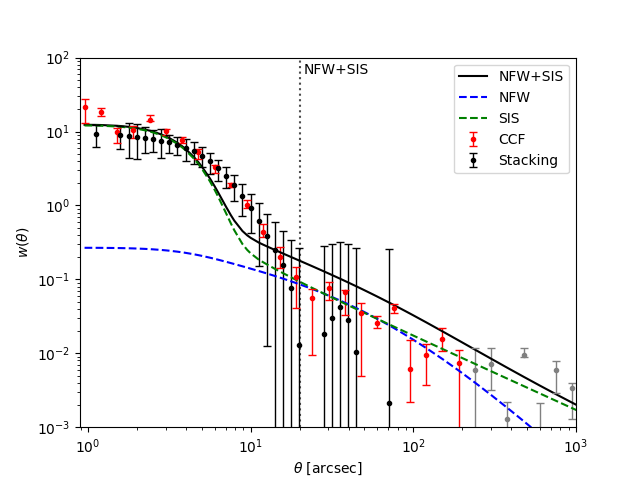}
 \includegraphics[width=0.49\textwidth]{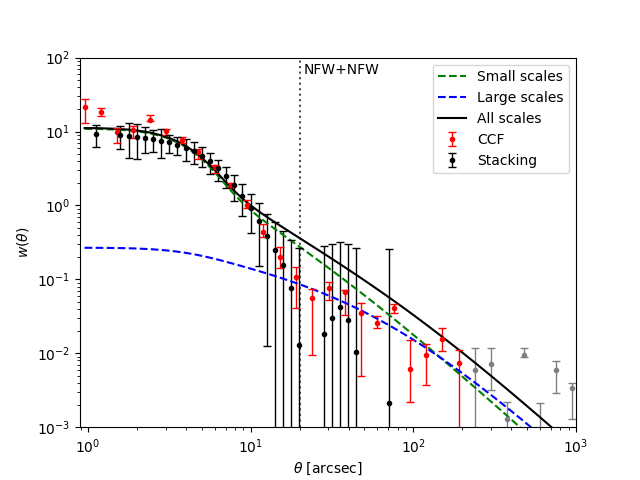}
 \caption{Stacking data (black points) and those of the cross-correlation function (red points) computed using the galaxies acting as lenses as in \cite{GON17}. On the left, the black line represents the NFW+SIS fits, where the corresponding SIS (green dashed line) and NFW (blue dashed line) fits are also shown separately. In the same way, on the right, the NFW+NFW profile is represented with the black line. In this case, the green and blue dashed lines correspond to the fits to the points at the small scales only and large scale only, respectively. The grey dashed vertical line at $\sim 20$ arcsec represents the separation between small and large scales. Grey points are considered outliers and are not taken into account for the analysis.}
 \label{fig:glx_xcorr}
\end{figure*}

As in the case of QSOs, with the stacking technique (black points in Fig. \ref{fig:glx_xcorr}) we obtain large error bars for the outer region that limit our measurements above $50$ arcsec. We therefore use the CCF measurements (red points) to better constrain the outer fit. The grey points are those excluded from the analysis.
There is good agreement between the measurements obtained with the stacking and the traditional CCF approaches, especially at the smallest scales (which cannot be fully appreciated in our analysis of QSOs) even if no smoothing is applied to the CCF. This is confirmation of the necessity to introduce the smoothing step in the stacking procedure and of the positional uncertainty value used.

We again perform a first fit to the cross-correlation data, combining a SIS and an NFW profile. The results are shown in the left panel of Fig. \ref{fig:glx_xcorr} with the black solid line. The dashed red and grey lines show the parts of the fit corresponding to the NFW and the SIS, respectively.

\begin{table}[ht]
\begin{tabular}{ccccc}
\hline \hline
                  & \small$log_{10}(M/M_{\odot})$ & \small$\Delta log_{10}(M/M_{\odot})$ & \small$C$ & \small$\Delta C$  \\ \hline
\small NFW$_{small}$     & $13.32$       & $0.08$  & $8.23$ & $0.77$ \\
\small SIS$_{small}$     & $12.784$ & $0.003$  & ---         & ---         \\
\small NFW$_{large-CCF}$ & 12.78       & 0.21  & 1.21 & 1.01 \\ \hline \hline
\end{tabular}
\caption{\label{table_glx}Derived parameter values for the galaxy sample for each profile and angular separation regime. The second and third columns show the mass and its error and the fourth and fifth columns contain the concentration and its error (only for the NFW profile).}
\end{table}

As can be seen in the left panel of Fig. \ref{fig:glx_xcorr}, the SIS+NFW does not produce a good fit to the measurements in both regions. The best fit is clearly below the measurements in the inner region, approximately between $5$ and $10$ arcsec, as in \cite{FER22}, and is above them in the outer region. The parameter values for the profiles are listed in Table \ref{table_glx}. The mass obtained for the inner SIS fit is $\log_{10}(M/M_{\odot})=12.784\pm0.003$ and the mass and concentration for the outer NFW fit are $\log_{10}(M/M_{\odot})=12.78\pm0.21$ and $C=1.21\pm1.01$, respectively. The correlation between the two parameters in this case is about 20\%.

The right panel of Fig. \ref{fig:glx_xcorr} shows the fit resulting from the combination of two independent NFW profiles. Overall, this fit is much better than the fit with the SIS+NFW profile. There is still disagreement with the data in the region of transition between the inner and outer part at about $20$ arcsec where the best fit is clearly above the measurements \citep[which is similar to the results derived from the galaxy cluster sample analysis by][]{FER22}. As listed in Table \ref{table_glx}, the value of the mass and concentration for the inner part are $\log_{10}(M/M_{\odot})=13.32\pm0.08$ and $C=8.23\pm0.77$, with a negligible correlation between the two parameters (6\%). The results for the outer part are the same as the SIS+NFW fit.

\subsection{Discussion}
\label{sec:discuss}

As described in sections \ref{sec:QSO} and \ref{sec:GLX}, the measured CCFs cannot be explained using a single mass density profile. This mirrors the conclusions of \citet{FER22} and those of previous works \citep[][]{BAU14,JOH07,OKA16}, but now with two additional independent samples with different kinds of lenses.

In this work, we follow the approach of \citet{FER22}, fitting the measured CCFs for both samples with a combination of two different mass density profiles (SIS+NFW and NFW+NFW). Although both combinations provide a reasonable fit, the SIS+NFW profile predicts a lower cross-correlation at angular separations between $\sim$5 and 10 arcsec, which is more obvious for the galaxy sample ($>5\sigma$ difference). This issue is produced by the SIS mass density profile that is naturally chosen to fit the inner region. The same problem was also identified by \citet{FER22} using the galaxy clusters sample. Therefore, we discard the NFW+SIS profile and focus on the NFW+NFW profile for the rest of the discussion.

In the inner region, the estimated lens average mass is almost the same for both samples, with $\log_{10}(M/M_{\odot})=13.51\pm0.04$ for the QSOs and $\log_{10}(M/M_{\odot})=13.32\pm0.08$ for the galaxies. These masses are in agreement with those expected for a typical large red galaxy (LRG): an effective halo mass of $M_{\text{eff}}=2.8-4.4\times10^{13}M_{\odot}$ is derived from the modelling of the LRG angular correlation function \citep{BLA08} and consistent masses are also estimated from the analysis of their large-scale redshift-space distortions $M=3.5^{+1.8}_{-1.4}\times10^{13}M_{\odot}$, \citep[][]{CAB09,BAU14}. The estimated masses are also in agreement with the central mass of M$_{\text{NFW}}=3-4\times10^{13}M_{\odot}$ derived for the galaxy clusters and this agreement is more or less independent of richness \citep[][]{FER22}. Therefore, we can conclude that our lenses are primarily LRGs, and that in the case of the galaxy clusters, these correspond to the brightest central galaxy (BCG), which has similar physical characteristics to the LRGs.

In the outer region, there is a clear difference between the two lens samples in their estimated lens
average mass. For the galaxies, the derived outer mass, $\log_{10}(M/M_{\odot})=$12.78$\pm$0.21, is about five times lower than the inner mass. We believe this is because the galaxies acting as lenses in this sample tend to be isolated, or at least they are the central galaxy of a small group with dwarf galaxies as satellites. However, for the QSO sample, the estimated outer mass, $\log_{10}(M/M_{\odot})=$13.44$\pm$0.17, is similar to the inner mass. Therefore, we can expect the QSOs to be the central galaxies of a small group of smaller galaxies. \cite{BON19} came to the same conclusion after analysing the same QSO lens sample.

We can also compare the outer mass derived for the QSO sample with the masses obtained by \citet[][]{FER22} for the different richness bins of the galaxy cluster sample (see Table \ref{table_cls}). The estimated outer mass for the QSOs is five to six times smaller than that obtained for the lowest richness bin (12--17 members). Considering that the outer mass increases with the richness, this result implies that each of the QSOs is the central galaxy of a group of galaxies with less than about ten members, which is in agreement with our previous conclusion.

With respect to concentration, in the outer region, all the lens samples (QSOs, galaxies, and also galaxy clusters) show similar very low concentration values, namely $C\lesssim$ 1.2. As already noted in \citet[][]{FER22}, these values are generally lower than the ones retrieved from the most common mass--concentration relationships \citep[e.g.][]{MAN08,DUT14,CHI18}. However, \citet[][]{BON19} estimate a concentration value of $C=3.5^{+0.5}_{-0.3}$ for the same QSO sample, which is in closer agreement with the mass--concentration relationships. The fact that the \citet[][]{BON19} results are obtained by analysing the whole range of angular separations, which is the most common procedure, implies that the overall concentrations could be overestimated due to the influence of the highly concentrated central regions of the halos. Moreover, for the inner regions, we find much higher concentrations for both lens samples: $C=6.85\pm0.34$ for the QSOs and $C=8.23\pm0.77$ for the galaxies. These values are higher than the one estimated by \cite{FER22} for the lowest richness bin of the galaxy cluster sample, namely $C=3.63$. 

It is very interesting that the findings of \citet{Luo22} for the satellite fraction can be extended to our case to explain the lower concentration values we obtain. In their work on dark-matter halos of luminous AGNs from galaxy--galaxy lensing, these latter authors modelled the halo occupation distribution (HOD) following \citet{MAN05} and set the satellite fraction as a free parameter. They find that a larger satellite fraction leads to an upturn at larger scales and flattens the overall profile which translates in a lower concentration value. Furthermore, at relatively larger scales, the excess surface density drops down towards the profile of less massive galaxies, which is explained by the fact that the satellite fraction increases causing a flattening of the profile. Analogously, for the samples adopted in this work, we treat the relatively large scales (which still belong to the one-halo term) separately from the small scales, which enhances the flattening effect, because in this region we are isolating the satellites and obtaining low concentration values. This flatting is what drove us to split the fit into two parts to better describe such different behaviours between the two angular scale ranges.

\begin{figure}[ht]
 \includegraphics[width=0.49\textwidth]{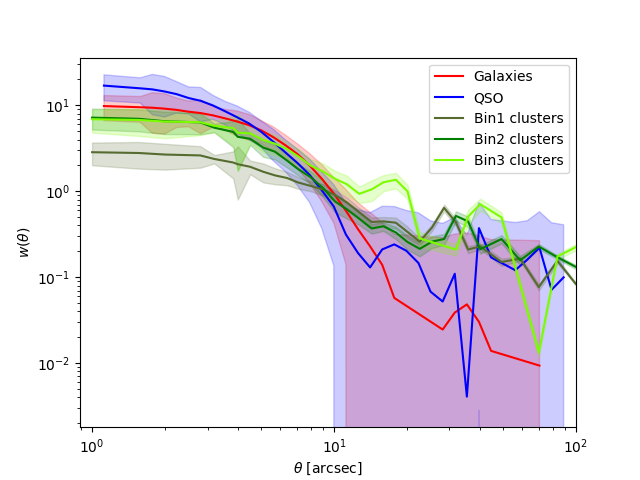} 
 \caption{Comparison of the different stacking results by means of continuous lines with their corresponding error bands. The stacking results obtained in this work are shown in blue for the case of QSOs and in red for the galaxy sample. Such results are compared to the cross-correlation data obtained in \cite{FER22}, where the galaxy clusters are acting as lenses: in particular, the BIN 1 case is shown in dark green, BIN 2 in green, and BIN 3 in light green, corresponding to the 12-17, 18-25, and 26-40 richness intervals, respectively.}
 \label{fig:lens_comparison_band}
 \end{figure}

 Figure \ref{fig:lens_comparison_band} shows the comparison between the cross-correlation measurements with stacking for the two different lens samples studied in this work. The data corresponding to the galaxy sample are represented with the red solid line, with the red coloured areas being the uncertainties, and those of the QSOs lens sample are shown with the blue solid line with their uncertainty represented by the blue coloured area. In addition, the cross-correlation obtained by \citet{FER22} for the three bins with the lowest richness is shown by the green solid lines (uncertainties are given by the green area).
This figure illustrates our previous conclusions: the galaxies can mostly be considered isolated halos, the QSO measurements are similar to those for galaxy clusters in the outer region (i.e. they are the main members of a smaller group of galaxies), and the BCG halos of the lowest richness are less concentrated with respect the QSOs and galaxies. This last point suggests an evolutionary and dynamical variation of the inner region concentration. The stacking results describe the average characteristics, and not the individual ones, and therefore the following discussion does not apply to a single galaxy but should be considered as a plausible average scenario.

We can use the galaxy sample results as a benchmark, considering that the inner mass is similar to the other cases and assuming that they are mainly isolated halos. On the other hand, the QSOs show a lower concentration and, by definition, have an AGN in their nuclei. Moreover, they show a lower signal with respect to the galaxy clusters between 10 and 30 arcsec ($\sim$70 --210 kpc at z=0.7), which corresponds to the outskirts of the host galaxy. Therefore, a plausible scenario to explain all these characteristics is that the host galaxy, as the BCG of a small group of galaxies, recently underwent an interaction with a surrounding satellite galaxy, which initiated the AGN episode.

On the other hand, the BCGs of the lowest richness bin of the galaxy clusters sample have similar mass but much lower concentration. In addition, the BCGs are mostly elliptical galaxies that are supposed to be the final product after the QSO phase. It is well established that the dynamical effects produced by an AGN affect the entire host galaxy (jets, outflows, star formation rate, etc.) and even its surroundings. Therefore, our results indicate that the BCGs of the smaller groups of galaxies could have less concentrated galaxy halos as a remnant of the QSO phase and AGN effect. 

Moreover, Fig. \ref{fig:lens_comparison_band} compares the cross-correlation function derived for the  galaxy clusters in the different richness bins. As already discussed in \citet{FER22}, the concentration increases with the mass or richness in our case. For the more massive galaxy clusters, which are older than the less massive ones, the BCG halos have had enough time to revert the AGN dynamical effects to a more concentrated mass density distribution due to gravity itself.

From this discussion, it is clear that a comparative analysis of mass density profiles estimated through the magnification bias for different kinds of lenses can be performed. Here, we demonstrated that this interesting and novel methodology can be used to study the average physical properties of the lenses and their surroundings and can potentially link these characteristics to the different galaxy evolution stages.

\begin{table*}[ht]
\centering
\begin{large}
\begin{tabular}{ccccccccc}

\hline 
              & \multicolumn{2}{c}{\begin{tabular}[c]{@{}c@{}}Bin 1\\ Richness {[}12-17{]}\end{tabular}} & \multicolumn{2}{c}{\begin{tabular}[c]{@{}c@{}}Bin 2\\ Richness {[}18-25{]}\end{tabular}} & \multicolumn{2}{c}{\begin{tabular}[c]{@{}c@{}}Bin 3\\ Richness {[}26-40{]}\end{tabular}} & \multicolumn{2}{c}{\begin{tabular}[c]{@{}c@{}}Total\\ Richness {[}12-220{]}\end{tabular}} \\ \hline \hline
              & {\small NFW$_{small}$}                               & {\small NFW$_{large}$ }                              & {\small NFW$_{small}$ }                              & {\small NFW$_{large}$ }                             & {\small NFW$_{small}$ }                              & {\small NFW$_{large}$ }                             & {\small NFW$_{small}$ }                            & {\small NFW$_{large}$  }                             \\ \hline
$log_{10}(M/M_\odot)$ & $13.58$                         & $13.76$                        & $13.37$                         & $13.89$                        & $13.86$                         & $14.05$                        & $13.61$                         & $13.85$                         \\
$C$           & $3.63$                                      & $0.74$                                     & $6.83$                                      & $0.39$                                     & $3.81$                                      & $1.00$                                     & $4.17$                                      & $1.72$                                      \\ \hline \hline

\end{tabular}
\end{large}
\caption{\label{table_cls} Estimated mass (first row) and concentration (second row) for the case of the galaxy clusters acting as lenses by means of the NFW+NFW profiles for small ($\lesssim 100 $ kpc) and large ($\gtrsim 100 $ kpc) scales derived by \citet[][]{FER22}. From left to right, the first three bins and the total case are summarised according to the best-fit mass density profile. For each of the four cases, the results  for small and large angular scales are shown.}
\end{table*}


\section{Conclusions}
\label{sec:concl}
In this work, we exploit the magnification bias ---a gravitational lensing effect produced on the SMGs observed by \textit{Herschel}--- in order to analyse the average mass density profiles of different types of lenses (QSOs, galaxies, and galaxy clusters). The magnification bias can be studied from the estimation of the CCF function, which we measure by stacking the background--foreground pairs with the Davis-Peebles estimator. 
One of the advantages of this approach is that positional uncertainties are easily taken into account, which are especially important at small angular scales. Moreover, with stacking data and including traditional CCF data at the largest angular scales, we can estimate the mass density profile for a large range of angular scales, $\sim 1-1000$ arcsec, considering both weak and strong lensing effects and including the inner part $< 100 kpc$ with high radial resolution. One the other hand, our results are average values obtained for samples and do not apply to single objects.

As already concluded by \citet{FER22}, the CCF measurements cannot be explained with a single mass density profile. Therefore, we combine two different mass density profiles: we use an NFW for the outer part and we fit the inner part both to a SIS and an NFW profile. 
The SIS profile always underestimates the data in the central part, especially in the galaxy sample case, and the two NFW profiles produce a better fit. For this reason, we focus our discussion on the NFW+NFW case.

In the inner region, the average mass of the lenses is similar for both QSOs and galaxy samples $log_{10}(M/M_{\odot})=$13.32 and 13.51, respectively, and agrees with the values found in literature for typical LRGs. Moreover, this value also agrees with the average central mass estimated by \cite{FER22} for galaxy clusters, which is almost richness independent.

In the outer region instead, there is a clear difference between the masses estimated for the two samples. In the galaxy case, the outer estimated mass is 100 times lower than the inner mass, which is probably due to the fact that these galaxies are isolated or are the central galaxies of galaxy groups with a few dwarf galaxies. On the other hand, the QSO sample provides outer estimated masses that are similar to those of the inner part. A comparison with the results of \cite{FER22} as a function of the richness of the galaxy clusters suggests that the QSOs are the central galaxies of galaxy groups \citep[as concluded in][]{BON19} with less than ten members.

Our results show that the concentrations of the different sources are very similar for the three samples (QSOs, galaxies, and galaxy clusters) in the outer region but are lower than the majority of values reported in literature, including those obtained by \cite{BON19} for the same QSO sample. This can be explained by the fact that these latter studies consider only one mass density profile for all angular scales, including the more concentrated central halo regions. This explanation is also supported by our higher concentration estimations in the small-scale region. This is also analogous to the findings of \citet{Luo22} for the satellite fraction: the increase in the satellite fraction at relatively large scales flattens the profile, which leads to a lower concentration. In our case, where the fit is done by isolating the data points at such scales, the effect on concentration is magnified.

Finally, the results from both angular scale regimes considered in our work suggest a scenario where the QSO host galaxies recently interacted with a close satellite galaxy, probably starting the AGN activity. Furthermore, the lower concentration values for the BCG halos in the galaxy clusters (mainly for the lowest richness bin) could be interpreted as the effects of the previous evolutionary activity produced during the earlier QSO and AGN phases acting on the surrounding environment. 
However, the galaxies in the galaxy sample and the BCGs in the more massive galaxy clusters, which are older than the less massive ones, had had enough time to weaken the AGN dynamical effects and drift back their halo into a more concentrated mass density distribution due to gravity itself.

\begin{acknowledgements}
We would like to warmly acknowledge the anonymous referee for the very useful comments that largely improved the work.
DC, JGN, LB, MMC, JMC acknowledge the PID2021-125630NB-I00 project funded by MCIN/AEI/10.13039/501100011033 / FEDER, UE.
MMC acknowledges PAPI-20-PF-23 (Universidad de Oviedo).\\

The \textit{Herschel}-ATLAS is a project with \textit{Herschel}, which is an ESA space observatory with science instruments provided by European-led Principal Investigator consortia and with important participation from NASA. The H-ATLAS web-site is http://www.h-atlas.org. GAMA is a joint European-Australasian project based around a spectroscopic campaign using the Anglo-Australian Telescope.\\

This research has made use of the python packages \texttt{ipython} \citep{ipython}, \texttt{matplotlib} \citep{matplotlib} and \texttt{Scipy} \citep{scipy}.
\end{acknowledgements}

\bibliographystyle{aa} 
\bibliography{Stack} 

\begin{thebibliography}{85}
\expandafter\ifx\csname natexlab\endcsname\relax\def\natexlab#1{#1}\fi

\bibitem[{{Allen} {et~al.}(2011){Allen}, {Evrard}, \& {Mantz}}]{ALL11}
{Allen}, S.~W., {Evrard}, A.~E., \& {Mantz}, A.~B. 2011, \araa, 49, 409

\bibitem[{{Baldry} {et~al.}(2014){Baldry}, {Alpaslan}, {Bauer}, {Bland
  -Hawthorn}, {Brough}, {Cluver}, {Croom}, {Davies}, {Driver}, {Gunawardhana},
  {Holwerda}, {Hopkins}, {Kelvin}, {Liske}, {L{\'o}pez-S{\'a}nchez}, {Loveday},
  {Norberg}, {Peacock}, {Robotham}, \& {Taylor}}]{BAL14}
{Baldry}, I.~K., {Alpaslan}, M., {Bauer}, A.~E., {et~al.} 2014, \mnras, 441,
  2440

\bibitem[{{Baldry} {et~al.}(2010){Baldry}, {Robotham}, {Hill}, {Driver},
  {Liske}, {Norberg}, {Bamford}, {Hopkins}, {Loveday}, {Peacock}, {Cameron},
  {Croom}, {Cross}, {Doyle}, {Dye}, {Frenk}, {Jones}, {van Kampen}, {Kelvin},
  {Nichol}, {Parkinson}, {Popescu}, {Prescott}, {Sharp}, {Sutherland},
  {Thomas}, \& {Tuffs}}]{BAL10}
{Baldry}, I.~K., {Robotham}, A.~S.~G., {Hill}, D.~T., {et~al.} 2010, \mnras,
  404, 86

\bibitem[{{Bartelmann} \& {Schneider}(1994)}]{Bar93}
{Bartelmann}, M. \& {Schneider}, P. 1994, \aap, 284, 1

\bibitem[{{Bartelmann} \& {Schneider}(2001)}]{Bar01}
{Bartelmann}, M. \& {Schneider}, P. 2001, Phys. Rep., 340, 291

\bibitem[{{Bauer} {et~al.}(2014){Bauer}, {Gazta{\~n}aga}, {Mart{\'\i}}, \&
  {Miquel}}]{BAU14}
{Bauer}, A.~H., {Gazta{\~n}aga}, E., {Mart{\'\i}}, P., \& {Miquel}, R. 2014,
  \mnras, 440, 3701

\bibitem[{{B{\'e}thermin} {et~al.}(2012){B{\'e}thermin}, {Le Floc'h}, {Ilbert},
  {Conley}, {Lagache}, {Amblard}, {Arumugam}, {Aussel}, {Berta}, {Bock},
  {Boselli}, {Buat}, {Casey}, {Castro-Rodr{\'{\i}}guez}, {Cava}, {Clements},
  {Cooray}, {Dowell}, {Eales}, {Farrah}, {Franceschini}, {Glenn}, {Griffin},
  {Hatziminaoglou}, {Heinis}, {Ibar}, {Ivison}, {Kartaltepe}, {Levenson},
  {Magdis}, {Marchetti}, {Marsden}, {Nguyen}, {O'Halloran}, {Oliver}, {Omont},
  {Page}, {Panuzzo}, {Papageorgiou}, {Pearson}, {P{\'e}rez-Fournon}, {Pohlen},
  {Rigopoulou}, {Roseboom}, {Rowan-Robinson}, {Salvato}, {Schulz}, {Scott},
  {Seymour}, {Shupe}, {Smith}, {Symeonidis}, {Trichas}, {Tugwell}, {Vaccari},
  {Valtchanov}, {Vieira}, {Viero}, {Wang}, {Xu}, \& {Zemcov}}]{Bet12}
{B{\'e}thermin}, M., {Le Floc'h}, E., {Ilbert}, O., {et~al.} 2012, \aap, 542,
  A58

\bibitem[{{Bianchini} {et~al.}(2019){Bianchini}, {Fabbian}, {Lapi},
  {Gonzalez-Nuevo}, {Gilli}, \& {Baccigalupi}}]{Bia19}
{Bianchini}, F., {Fabbian}, G., {Lapi}, A., {et~al.} 2019, \apj, 871, 136

\bibitem[{{Bianchini} \& {Reichardt}(2018)}]{Bia18}
{Bianchini}, F. \& {Reichardt}, C.~L. 2018, \apj, 862, 81

\bibitem[{{Blain} {et~al.}(1999){Blain}, {Smail}, {Ivison}, \& {Kneib}}]{BLA99}
{Blain}, A.~W., {Smail}, I., {Ivison}, R.~J., \& {Kneib}, J.~P. 1999, \mnras,
  302, 632

\bibitem[{Blake {et~al.}(2008)Blake, Collister, \& Lahav}]{BLA08}
Blake, C., Collister, A., \& Lahav, O. 2008, Monthly Notices of the Royal
  Astronomical Society, 385, 1257

\bibitem[{{Bonavera} {et~al.}(2021{\natexlab{a}}){Bonavera}, {Cueli}, \&
  {Gonzalez-Nuevo}}]{BON21P}
{Bonavera}, L., {Cueli}, M.~M., \& {Gonzalez-Nuevo}, J. 2021{\natexlab{a}},
  arXiv e-prints, arXiv:2112.02959

\bibitem[{{Bonavera} {et~al.}(2021{\natexlab{b}}){Bonavera}, {Cueli},
  {Gonz{\'a}lez-Nuevo}, {Ronconi}, {Migliaccio}, {Lapi}, {Casas}, \&
  {Crespo}}]{BON21}
{Bonavera}, L., {Cueli}, M.~M., {Gonz{\'a}lez-Nuevo}, J., {et~al.}
  2021{\natexlab{b}}, \aap, 656, A99

\bibitem[{{Bonavera} {et~al.}(2017{\natexlab{a}}){Bonavera},
  {Gonz{\'a}lez-Nuevo}, {Arg{\"u}eso}, \& {Toffolatti}}]{BON17a}
{Bonavera}, L., {Gonz{\'a}lez-Nuevo}, J., {Arg{\"u}eso}, F., \& {Toffolatti},
  L. 2017{\natexlab{a}}, \mnras, 469, 2401

\bibitem[{{Bonavera} {et~al.}(2020){Bonavera}, {Gonz{\'a}lez-Nuevo}, {Cueli},
  {Ronconi}, {Migliaccio}, {Dunne}, {Lapi}, {Maddox}, \& {Negrello}}]{BON20}
{Bonavera}, L., {Gonz{\'a}lez-Nuevo}, J., {Cueli}, M.~M., {et~al.} 2020, \aap,
  639, A128

\bibitem[{{Bonavera} {et~al.}(2017{\natexlab{b}}){Bonavera},
  {Gonz{\'a}lez-Nuevo}, {De Marco}, {Arg{\"u}eso}, \& {Toffolatti}}]{BON17b}
{Bonavera}, L., {Gonz{\'a}lez-Nuevo}, J., {De Marco}, B., {Arg{\"u}eso}, F., \&
  {Toffolatti}, L. 2017{\natexlab{b}}, \mnras, 472, 628

\bibitem[{{Bonavera} {et~al.}(2019){Bonavera}, {Gonz{\'a}lez-Nuevo},
  {Su{\'a}rez G{\'o}mez}, {Lapi}, {Bianchini}, {Negrello}, {D{\'\i}ez Alonso},
  {Santos}, \& {de Cos Juez}}]{BON19}
{Bonavera}, L., {Gonz{\'a}lez-Nuevo}, J., {Su{\'a}rez G{\'o}mez}, S.~L.,
  {et~al.} 2019, \jcap, 2019, 021

\bibitem[{{Bourne} {et~al.}(2016){Bourne}, {Dunne}, {Maddox}, {Dye},
  {Furlanetto}, {Hoyos}, {Smith}, {Eales}, {Smith}, {Valiante}, {Alpaslan},
  {Andrae}, {Baldry}, {Cluver}, {Cooray}, {Driver}, {Dunlop}, {Grootes},
  {Ivison}, {Jarrett}, {Liske}, {Madore}, {Popescu}, {Robotham}, {Rowland s},
  {Seibert}, {Thompson}, {Tuffs}, {Viaene}, \& {Wright}}]{BOU16}
{Bourne}, N., {Dunne}, L., {Maddox}, S.~J., {et~al.} 2016, \mnras, 462, 1714

\bibitem[{{Butcher} \& {Oemler}(1978)}]{BUT78}
{Butcher}, H. \& {Oemler}, A., J. 1978, \apj, 226, 559

\bibitem[{{Butcher} \& {Oemler}(1984)}]{BUT84}
{Butcher}, H. \& {Oemler}, A., J. 1984, \apj, 285, 426

\bibitem[{{Cabr{\'e}} \& {Gazta{\~n}aga}(2009)}]{CAB09}
{Cabr{\'e}}, A. \& {Gazta{\~n}aga}, E. 2009, \mnras, 396, 1119

\bibitem[{Child {et~al.}(2018)Child, Habib, Heitmann, Frontiere, Finkel, Pope,
  \& Morozov}]{CHI18}
Child, H.~L., Habib, S., Heitmann, K., {et~al.} 2018, The Astrophysical
  Journal, 859, 55

\bibitem[{{Courbin} {et~al.}(2012){Courbin}, {Faure}, {Djorgovski},
  {R{\'e}rat}, {Tewes}, {Meylan}, {Stern}, {Mahabal}, {Boroson}, {Dheeraj}, \&
  {Sluse}}]{COU12}
{Courbin}, F., {Faure}, C., {Djorgovski}, S.~G., {et~al.} 2012, \aap, 540, A36

\bibitem[{{Courbin} {et~al.}(2010){Courbin}, {Tewes}, {Djorgovski}, {Sluse},
  {Mahabal}, {R{\'e}rat}, \& {Meylan}}]{COU10}
{Courbin}, F., {Tewes}, M., {Djorgovski}, S.~G., {et~al.} 2010, \aap, 516, L12

\bibitem[{{Cueli} {et~al.}(2022){Cueli}, {Bonavera}, {Gonz{\'a}lez-Nuevo},
  {Crespo}, {Casas}, \& {Lapi}}]{CUE22}
{Cueli}, M.~M., {Bonavera}, L., {Gonz{\'a}lez-Nuevo}, J., {et~al.} 2022, \aap,
  662, A44

\bibitem[{{Cueli} {et~al.}(2021){Cueli}, {Bonavera}, {Gonz{\'a}lez-Nuevo}, \&
  {Lapi}}]{CUE21}
{Cueli}, M.~M., {Bonavera}, L., {Gonz{\'a}lez-Nuevo}, J., \& {Lapi}, A. 2021,
  \aap, 645, A126

\bibitem[{{Danielson} {et~al.}(2017){Danielson}, {Swinbank}, {Smail},
  {Simpson}, {Casey}, {Chapman}, {da Cunha}, {Hodge}, {Walter}, {Wardlow},
  {Alexander}, {Brandt}, {de Breuck}, {Coppin}, {Dannerbauer}, {Dickinson},
  {Edge}, {Gawiser}, {Ivison}, {Karim}, {Kovacs}, {Lutz}, {Menten},
  {Schinnerer}, {Wei{\ss}}, \& {van der Werf}}]{Dan17}
{Danielson}, A.~L.~R., {Swinbank}, A.~M., {Smail}, I., {et~al.} 2017, \apj,
  840, 78

\bibitem[{Davis \& Peebles(1983)}]{Dav83}
Davis, M. \& Peebles, P. 1983, The Astrophysical Journal, 267, 465

\bibitem[{{Dawson} {et~al.}(2013){Dawson}, {Schlegel}, {Ahn}, {Anderson},
  {Auonaveraourg}, {Bailey}, {Barkhouser}, {Bautista}, {Beifiori}, {Berlind},
  {Bhardwaj}, {Bizyaev}, {Blake}, {Blanton}, {Blomqvist}, {Bolton}, {Borde},
  {Bovy}, {Brandt}, {Brewington}, {Brinkmann}, {Brown}, {Brownstein}, {Bundy},
  {Busca}, {Carithers}, {Carnero}, {Carr}, {Chen}, {Comparat}, {Connolly},
  {Cope}, {Croft}, {Cuesta}, {da Costa}, {Davenport}, {Delubac}, {de Putter},
  {Dhital}, {Ealet}, {Ebelke}, {Eisenstein}, {Escoffier}, {Fan}, {Filiz Ak},
  {Finley}, {Font-Ribera}, {G{\'e}nova-Santos}, {Gunn}, {Guo}, {Haggard},
  {Hall}, {Hamilton}, {Harris}, {Harris}, {Ho}, {Hogg}, {Holder}, {Honscheid},
  {Huehnerhoff}, {Jordan}, {Jordan}, {Kauffmann}, {Kazin}, {Kirkby}, {Klaene},
  {Kneib}, {Le Goff}, {Lee}, {Long}, {Loomis}, {Lundgren}, {Lupton}, {Maia},
  {Makler}, {Malanushenko}, {Malanushenko}, {Mandelbaum}, {Manera}, {Maraston},
  {Margala}, {Masters}, {McBride}, {McDonald}, {McGreer}, {McMahon}, {Mena},
  {Miralda-Escud{\'e}}, {Montero-Dorta}, {Montesano}, {Muna}, {Myers},
  {Naugle}, {Nichol}, {Noterdaeme}, {Nuza}, {Olmstead}, {Oravetz}, {Oravetz},
  {Owen}, {Padmanabhan}, {Palanque-Delabrouille}, {Pan}, {Parejko},
  {P{\^a}ris}, {Percival}, {P{\'e}rez-Fournon}, {P{\'e}rez-R{\`a}fols},
  {Petitjean}, {Pfaffenberger}, {Pforr}, {Pieri}, {Prada}, {Price-Whelan},
  {Raddick}, {Rebolo}, {Rich}, {Richards}, {Rockosi}, {Roe}, {Ross}, {Ross},
  {Rossi}, {Rubi{\~n}o-Martin}, {Samushia}, {S{\'a}nchez}, {Sayres}, {Schmidt},
  {Schneider}, {Sc{\'o}ccola}, {Seo}, {Shelden}, {Sheldon}, {Shen}, {Shu},
  {Slosar}, {Smee}, {Snedden}, {Stauffer}, {Steele}, {Strauss}, {Streblyanska},
  {Suzuki}, {Swanson}, {Tal}, {Tanaka}, {Thomas}, {Tinker}, {Tojeiro},
  {Tremonti}, {Vargas Maga{\~n}a}, {Verde}, {Viel}, {Wake}, {Watson}, {Weaver},
  {Weinberg}, {Weiner}, {West}, {White}, {Wood-Vasey}, {Yeche}, {Zehavi},
  {Zhao}, \& {Zheng}}]{DAW13}
{Dawson}, K.~S., {Schlegel}, D.~J., {Ahn}, C.~P., {et~al.} 2013, \aj, 145, 10

\bibitem[{{Dole} {et~al.}(2006){Dole}, {Lagache}, {Puget}, {Caputi},
  {Fern{\'a}ndez-Conde}, {Le Floc'h}, {Papovich}, {P{\'e}rez-Gonz{\'a}lez},
  {Rieke}, \& {Blaylock}}]{Dol06}
{Dole}, H., {Lagache}, G., {Puget}, J.-L., {et~al.} 2006, \aap, 451, 417

\bibitem[{{Dressler}(1980)}]{DRE80}
{Dressler}, A. 1980, \apj, 236, 351

\bibitem[{{Driver} {et~al.}(2011){Driver}, {Hill}, {Kelvin}, {Robotham},
  {Liske}, {Norberg}, {Baldry}, {Bamford}, {Hopkins}, {Loveday}, {Peacock},
  {Andrae}, {Bland -Hawthorn}, {Brough}, {Brown}, {Cameron}, {Ching},
  {Colless}, {Conselice}, {Croom}, {Cross}, {de Propris}, {Dye}, {Drinkwater},
  {Ellis}, {Graham}, {Grootes}, {Gunawardhana}, {Jones}, {van Kampen},
  {Maraston}, {Nichol}, {Parkinson}, {Phillipps}, {Pimbblet}, {Popescu},
  {Prescott}, {Roseboom}, {Sadler}, {Sansom}, {Sharp}, {Smith}, {Taylor},
  {Thomas}, {Tuffs}, {Wijesinghe}, {Dunne}, {Frenk}, {Jarvis}, {Madore},
  {Meyer}, {Seibert}, {Staveley-Smith}, {Sutherland}, \& {Warren}}]{DRI11}
{Driver}, S.~P., {Hill}, D.~T., {Kelvin}, L.~S., {et~al.} 2011, \mnras, 413,
  971

\bibitem[{{Dunne} {et~al.}(2020){Dunne}, {Bonavera}, {Gonzalez-Nuevo},
  {Maddox}, \& {Vlahakis}}]{DUN20}
{Dunne}, L., {Bonavera}, L., {Gonzalez-Nuevo}, J., {Maddox}, S.~J., \&
  {Vlahakis}, C. 2020, \mnras, 498, 4635

\bibitem[{{Dutton} \& {Macci{\`o}}(2014)}]{DUT14}
{Dutton}, A.~A. \& {Macci{\`o}}, A.~V. 2014, \mnras, 441, 3359

\bibitem[{{Eales} {et~al.}(2010){Eales}, {Dunne}, {Clements}, {Cooray}, {De
  Zotti}, {Dye}, {Ivison}, {Jarvis}, {Lagache}, {Maddox}, {Negrello},
  {Serjeant}, {Thompson}, {Van Kampen}, {Amblard}, {Andreani}, {Baes},
  {Beelen}, {Bendo}, {Benford}, {Bertoldi}, {Bock}, {Bonfield}, {Boselli},
  {Bridge}, {Buat}, {Burgarella}, {Carlberg}, {Cava}, {Chanial}, {Charlot},
  {Christopher}, {Coles}, {Cortese}, {Dariush}, {da Cunha}, {Dalton}, {Danese},
  {Dannerbauer}, {Driver}, {Dunlop}, {Fan}, {Farrah}, {Frayer}, {Frenk},
  {Geach}, {Gardner}, {Gomez}, {Gonz{\'a}lez-Nuevo}, {Gonz{\'a}lez-Solares},
  {Griffin}, {Hardcastle}, {Hatziminaoglou}, {Herranz}, {Hughes}, {Ibar},
  {Jeong}, {Lacey}, {Lapi}, {Lawrence}, {Lee}, {Leeuw}, {Liske},
  {L{\'o}pez-Caniego}, {M{\"u}ller}, {Nandra}, {Panuzzo}, {Papageorgiou},
  {Patanchon}, {Peacock}, {Pearson}, {Phillipps}, {Pohlen}, {Popescu},
  {Rawlings}, {Rigby}, {Rigopoulou}, {Robotham}, {Rodighiero}, {Sansom},
  {Schulz}, {Scott}, {Smith}, {Sibthorpe}, {Smail}, {Stevens}, {Sutherland},
  {Takeuchi}, {Tedds}, {Temi}, {Tuffs}, {Trichas}, {Vaccari}, {Valtchanov},
  {van der Werf}, {Verma}, {Vieria}, {Vlahakis}, \& {White}}]{EAL10}
{Eales}, S., {Dunne}, L., {Clements}, D., {et~al.} 2010, \pasp, 122, 499

\bibitem[{{Fernandez} {et~al.}(2022){Fernandez}, {Cueli}, {Gonz{\'a}lez-Nuevo},
  {Bonavera}, {Crespo}, {Casas}, \& {Lapi}}]{FER22}
{Fernandez}, L., {Cueli}, M.~M., {Gonz{\'a}lez-Nuevo}, J., {et~al.} 2022, \aap,
  658, A19

\bibitem[{{Gonz{\'a}lez-Nuevo} {et~al.}(2021){Gonz{\'a}lez-Nuevo}, {Cueli},
  {Bonavera}, {Lapi}, {Migliaccio}, {Arg{\"u}eso}, \& {Toffolatti}}]{GON21}
{Gonz{\'a}lez-Nuevo}, J., {Cueli}, M.~M., {Bonavera}, L., {et~al.} 2021, \aap,
  646, A152

\bibitem[{{Gonz{\'a}lez-Nuevo} {et~al.}(2017){Gonz{\'a}lez-Nuevo}, {Lapi},
  {Bonavera}, {Danese}, {de Zotti}, {Negrello}, {Bourne}, {Cooray}, {Dunne},
  {Dye}, {Eales}, {Furlanetto}, {Ivison}, {Loveday}, {Maddox}, {Smith}, \&
  {Valiante}}]{GON17}
{Gonz{\'a}lez-Nuevo}, J., {Lapi}, A., {Bonavera}, L., {et~al.} 2017, \jcap,
  2017, 024

\bibitem[{{Gonz{\'a}lez-Nuevo} {et~al.}(2012){Gonz{\'a}lez-Nuevo}, {Lapi},
  {Fleuren}, {Bressan}, {Danese}, {De Zotti}, {Negrello}, {Cai}, {Fan},
  {Sutherland}, {Baes}, {Baker}, {Clements}, {Cooray}, {Dannerbauer}, {Dunne},
  {Dye}, {Eales}, {Frayer}, {Harris}, {Ivison}, {Jarvis}, {Micha{\l}owski},
  {L{\'o}pez-Caniego}, {Rodighiero}, {Rowlands}, {Serjeant}, {Scott}, {van der
  Werf}, {Auld}, {Buttiglione}, {Cava}, {Dariush}, {Fritz}, {Hopwood}, {Ibar},
  {Maddox}, {Pascale}, {Pohlen}, {Rigby}, {Smith}, \& {Temi}}]{GON12}
{Gonz{\'a}lez-Nuevo}, J., {Lapi}, A., {Fleuren}, S., {et~al.} 2012, \apj, 749,
  65

\bibitem[{{Gonz{\'a}lez-Nuevo} {et~al.}(2014){Gonz{\'a}lez-Nuevo}, {Lapi},
  {Negrello}, {Danese}, {De Zotti}, {Amber}, {Baes}, {Bland -Hawthorn},
  {Bourne}, {Brough}, {Bussmann}, {Cai}, {Cooray}, {Driver}, {Dunne}, {Dye},
  {Eales}, {Ibar}, {Ivison}, {Liske}, {Loveday}, {Maddox}, {Micha{\l}owski},
  {Robotham}, {Scott}, {Smith}, {Valiante}, \& {Xia}}]{GON14}
{Gonz{\'a}lez-Nuevo}, J., {Lapi}, A., {Negrello}, M., {et~al.} 2014, \mnras,
  442, 2680

\bibitem[{{Goto} {et~al.}(2003){Goto}, {Yamauchi}, {Fujita}, {Okamura},
  {Sekiguchi}, {Smail}, {Bernardi}, \& {Gomez}}]{GOT03}
{Goto}, T., {Yamauchi}, C., {Fujita}, Y., {et~al.} 2003, \mnras, 346, 601

\bibitem[{{Griffin} {et~al.}(2010){Griffin}, {Abergel}, {Abreu}, {Ade},
  {Andr{\'e}}, {Augueres}, {Babbedge}, {Bae}, {Baillie}, {Baluteau}, {Barlow},
  {Bendo}, {Benielli}, {Bock}, {Bonhomme}, {Brisbin}, {Brockley-Blatt},
  {Caldwell}, {Cara}, {Castro-Rodriguez}, {Cerulli}, {Chanial}, {Chen},
  {Clark}, {Clements}, {Clerc}, {Coker}, {Communal}, {Conversi}, {Cox},
  {Crumb}, {Cunningham}, {Daly}, {Davis}, {de Antoni}, {Delderfield}, {Devin},
  {di Giorgio}, {Didschuns}, {Dohlen}, {Donati}, {Dowell}, {Dowell}, {Duband},
  {Dumaye}, {Emery}, {Ferlet}, {Ferrand}, {Fontignie}, {Fox}, {Franceschini},
  {Frerking}, {Fulton}, {Garcia}, {Gastaud}, {Gear}, {Glenn}, {Goizel},
  {Griffin}, {Grundy}, {Guest}, {Guillemet}, {Hargrave}, {Harwit}, {Hastings},
  {Hatziminaoglou}, {Herman}, {Hinde}, {Hristov}, {Huang}, {Imhof}, {Isaak},
  {Israelsson}, {Ivison}, {Jennings}, {Kiernan}, {King}, {Lange}, {Latter},
  {Laurent}, {Laurent}, {Leeks}, {Lellouch}, {Levenson}, {Li}, {Li},
  {Lilienthal}, {Lim}, {Liu}, {Lu}, {Madden}, {Mainetti}, {Marliani}, {McKay},
  {Mercier}, {Molinari}, {Morris}, {Moseley}, {Mulder}, {Mur}, {Naylor},
  {Nguyen}, {O'Halloran}, {Oliver}, {Olofsson}, {Olofsson}, {Orfei}, {Page},
  {Pain}, {Panuzzo}, {Papageorgiou}, {Parks}, {Parr-Burman}, {Pearce},
  {Pearson}, {P{\'e}rez-Fournon}, {Pinsard}, {Pisano}, {Podosek}, {Pohlen},
  {Polehampton}, {Pouliquen}, {Rigopoulou}, {Rizzo}, {Roseboom}, {Roussel},
  {Rowan-Robinson}, {Rownd}, {Saraceno}, {Sauvage}, {Savage}, {Savini},
  {Sawyer}, {Scharmberg}, {Schmitt}, {Schneider}, {Schulz}, {Schwartz},
  {Shafer}, {Shupe}, {Sibthorpe}, {Sidher}, {Smith}, {Smith}, {Smith},
  {Spencer}, {Stobie}, {Sudiwala}, {Sukhatme}, {Surace}, {Stevens}, {Swinyard},
  {Trichas}, {Tourette}, {Triou}, {Tseng}, {Tucker}, {Turner}, {Vaccari},
  {Valtchanov}, {Vigroux}, {Virique}, {Voellmer}, {Walker}, {Ward}, {Waskett},
  {Weilert}, {Wesson}, {White}, {Whitehouse}, {Wilson}, {Winter}, {Woodcraft},
  {Wright}, {Xu}, {Zavagno}, {Zemcov}, {Zhang}, \& {Zonca}}]{GRI10}
{Griffin}, M.~J., {Abergel}, A., {Abreu}, A., {et~al.} 2010, \aap, 518, L3

\bibitem[{{Harvey} \& {Courbin}(2015)}]{Har15}
{Harvey}, D. \& {Courbin}, F. 2015, \mnras, 451, L95

\bibitem[{{Hildebrandt} {et~al.}(2013){Hildebrandt}, {van Waerbeke}, {Scott},
  {B{\'e}thermin}, {Bock}, {Clements}, {Conley}, {Cooray}, {Dunlop}, {Eales},
  {Erben}, {Farrah}, {Franceschini}, {Glenn}, {Halpern}, {Heinis}, {Ivison},
  {Marsden}, {Oliver}, {Page}, {P{\'e}rez-Fournon}, {Smith}, {Rowan-Robinson},
  {Valtchanov}, {van der Burg}, {Vieira}, {Viero}, \& {Wang}}]{Hil13}
{Hildebrandt}, H., {van Waerbeke}, L., {Scott}, D., {et~al.} 2013, \mnras, 429,
  3230

\bibitem[{Hunter(2007)}]{matplotlib}
Hunter, J.~D. 2007, Computing In Science \& Engineering, 9, 90

\bibitem[{{Ibar} {et~al.}(2010){Ibar}, {Ivison}, {Cava}, {Rodighiero},
  {Buttiglione}, {Temi}, {Frayer}, {Fritz}, {Leeuw}, {Baes}, {Rigby}, {Verma},
  {Serjeant}, {M{\"u}ller}, {Auld}, {Dariush}, {Dunne}, {Eales}, {Maddox},
  {Panuzzo}, {Pascale}, {Pohlen}, {Smith}, {de Zotti}, {Vaccari}, {Hopwood},
  {Cooray}, {Burgarella}, \& {Jarvis}}]{IBA10}
{Ibar}, E., {Ivison}, R.~J., {Cava}, A., {et~al.} 2010, \mnras, 409, 38

\bibitem[{{Ivison} {et~al.}(2016){Ivison}, {Lewis}, {Weiss}, {Arumugam},
  {Simpson}, {Holland}, {Maddox}, {Dunne}, {Valiante}, {van der Werf}, {Omont},
  {Dannerbauer}, {Smail}, {Bertoldi}, {Bremer}, {Bussmann}, {Cai}, {Clements},
  {Cooray}, {De Zotti}, {Eales}, {Fuller}, {Gonzalez-Nuevo}, {Ibar},
  {Negrello}, {Oteo}, {P{\'e}rez-Fournon}, {Riechers}, {Stevens}, {Swinbank},
  \& {Wardlow}}]{IVI16}
{Ivison}, R.~J., {Lewis}, A.~J.~R., {Weiss}, A., {et~al.} 2016, \apj, 832, 78

\bibitem[{{Ivison} {et~al.}(2010){Ivison}, {Swinbank}, {Swinyard}, {Smail},
  {Pearson}, {Rigopoulou}, {Polehampton}, {Baluteau}, {Barlow}, {Blain},
  {Bock}, {Clements}, {Coppin}, {Cooray}, {Danielson}, {Dwek}, {Edge},
  {Franceschini}, {Fulton}, {Glenn}, {Griffin}, {Isaak}, {Leeks}, {Lim},
  {Naylor}, {Oliver}, {Page}, {P{\'e}rez Fournon}, {Rowan-Robinson}, {Savini},
  {Scott}, {Spencer}, {Valtchanov}, {Vigroux}, \& {Wright}}]{IVI10}
{Ivison}, R.~J., {Swinbank}, A.~M., {Swinyard}, B., {et~al.} 2010, \aap, 518,
  L35

\bibitem[{{Johnston} {et~al.}(2007){Johnston}, {Sheldon}, {Wechsler}, {Rozo},
  {Koester}, {Frieman}, {McKay}, {Evrard}, {Becker}, \& {Annis}}]{JOH07}
{Johnston}, D.~E., {Sheldon}, E.~S., {Wechsler}, R.~H., {et~al.} 2007, arXiv
  e-prints, arXiv:0709.1159

\bibitem[{Jones {et~al.}(2001)Jones, Oliphant, Peterson, {et~al.}}]{scipy}
Jones, E., Oliphant, T., Peterson, P., {et~al.} 2001, {SciPy}: Open source
  scientific tools for {Python}

\bibitem[{{Lapi} {et~al.}(2011){Lapi}, {Gonz{\'a}lez-Nuevo}, {Fan}, {Bressan},
  {De Zotti}, {Danese}, {Negrello}, {Dunne}, {Eales}, {Maddox}, {Auld}, {Baes},
  {Bonfield}, {Buttiglione}, {Cava}, {Clements}, {Cooray}, {Dariush}, {Dye},
  {Fritz}, {Herranz}, {Hopwood}, {Ibar}, {Ivison}, {Jarvis}, {Kaviraj},
  {L{\'o}pez-Caniego}, {Massardi}, {Micha{\l}owski}, {Pascale}, {Pohlen},
  {Rigby}, {Rodighiero}, {Serjeant}, {Smith}, {Temi}, {Wardlow}, \& {van der
  Werf}}]{LAP11}
{Lapi}, A., {Gonz{\'a}lez-Nuevo}, J., {Fan}, L., {et~al.} 2011, \apj, 742, 24

\bibitem[{Liske {et~al.}(2015)Liske, Baldry, Driver, Tuffs, Alpaslan, Andrae,
  Brough, Cluver, Grootes, Gunawardhana, {et~al.}}]{LIS15}
Liske, J., Baldry, I.~K., Driver, S.~P., {et~al.} 2015, Monthly Notices of the
  Royal Astronomical Society, 452, 2087

\bibitem[{{Lopez} {et~al.}(2008){Lopez}, {Barrientos}, {Lira}, {Padilla},
  {Gilbank}, {Gladders}, {Maza}, {Tejos}, {Vidal}, \& {Yee}}]{LOP08}
{Lopez}, S., {Barrientos}, L.~F., {Lira}, P., {et~al.} 2008, \apj, 679, 1144

\bibitem[{{Luo} {et~al.}(2022){Luo}, {Silverman}, {More}, {Goulding},
  {Miyatake}, {Nishimichi}, {Hikage}, {Kawinwanichakij}, {Li}, {Li},
  {Medezinski}, {Oguri}, {Oogi}, \& {Sifon}}]{Luo22}
{Luo}, W., {Silverman}, J.~D., {More}, S., {et~al.} 2022, arXiv e-prints,
  arXiv:2204.03817

\bibitem[{{Maddox} {et~al.}(2018){Maddox}, {Valiante}, {Cigan}, {Dunne},
  {Eales}, {Smith}, {Dye}, {Furlanetto}, {Ibar}, {de Zotti}, {Millard},
  {Bourne}, {Gomez}, {Ivison}, {Scott}, \& {Valtchanov}}]{MAD18}
{Maddox}, S.~J., {Valiante}, E., {Cigan}, P., {et~al.} 2018, \apjs, 236, 30

\bibitem[{{Mandelbaum} {et~al.}(2009){Mandelbaum}, {Li}, {Kauffmann}, \&
  {White}}]{Man09}
{Mandelbaum}, R., {Li}, C., {Kauffmann}, G., \& {White}, S. D.~M. 2009, \mnras,
  393, 377

\bibitem[{{Mandelbaum} {et~al.}(2008){Mandelbaum}, {Seljak}, \&
  {Hirata}}]{MAN08}
{Mandelbaum}, R., {Seljak}, U., \& {Hirata}, C.~M. 2008, \jcap, 2008, 006

\bibitem[{{Mandelbaum} {et~al.}(2005){Mandelbaum}, {Tasitsiomi}, {Seljak},
  {Kravtsov}, \& {Wechsler}}]{MAN05}
{Mandelbaum}, R., {Tasitsiomi}, A., {Seljak}, U., {Kravtsov}, A.~V., \&
  {Wechsler}, R.~H. 2005, \mnras, 362, 1451

\bibitem[{{Marsden} {et~al.}(2009){Marsden}, {Ade}, {Bock}, {Chapin}, {Devlin},
  {Dicker}, {Griffin}, {Gundersen}, {Halpern}, {Hargrave}, {Hughes}, {Klein},
  {Mauskopf}, {Magnelli}, {Moncelsi}, {Netterfield}, {Ngo}, {Olmi}, {Pascale},
  {Patanchon}, {Rex}, {Scott}, {Semisch}, {Thomas}, {Truch}, {Tucker},
  {Tucker}, {Viero}, \& {Wiebe}}]{Mar09}
{Marsden}, G., {Ade}, P.~A.~R., {Bock}, J.~J., {et~al.} 2009, \apj, 707, 1729

\bibitem[{{M{\'e}nard} {et~al.}(2010){M{\'e}nard}, {Scranton}, {Fukugita}, \&
  {Richards}}]{Men10}
{M{\'e}nard}, B., {Scranton}, R., {Fukugita}, M., \& {Richards}, G. 2010,
  \mnras, 405, 1025

\bibitem[{{Myers} {et~al.}(2005){Myers}, {Outram}, {Shanks}, {Boyle}, {Croom},
  {Loaring}, {Miller}, \& {Smith}}]{MYE05}
{Myers}, A.~D., {Outram}, P.~J., {Shanks}, T., {et~al.} 2005, \mnras, 359, 741

\bibitem[{{Navarro} {et~al.}(1996){Navarro}, {Frenk}, \& {White}}]{NAV96}
{Navarro}, J.~F., {Frenk}, C.~S., \& {White}, S. D.~M. 1996, \apj, 462, 563

\bibitem[{{Oguri} {et~al.}(2006){Oguri}, {Inada}, {Pindor}, {Strauss},
  {Richards}, {Hennawi}, {Turner}, {Lupton}, {Schneider}, {Fukugita}, \&
  {Brinkmann}}]{Ogu06}
{Oguri}, M., {Inada}, N., {Pindor}, B., {et~al.} 2006, \aj, 132, 999

\bibitem[{{Oguri} {et~al.}(2008){Oguri}, {Inada}, {Strauss}, {Kochanek},
  {Richards}, {Schneider}, {Becker}, {Fukugita}, {Gregg}, {Hall}, {Hennawi},
  {Johnston}, {Kayo}, {Keeton}, {Pindor}, {Shin}, {Turner}, {White}, {York},
  {Anderson}, {Bahcall}, {Brunner}, {Burles}, {Castander}, {Chiu},
  {Clocchiatti}, {Eisenstein}, {Frieman}, {Kawano}, {Lupton}, {Morokuma},
  {Rix}, {Scranton}, \& {Sheldon}}]{Ogu08}
{Oguri}, M., {Inada}, N., {Strauss}, M.~A., {et~al.} 2008, \aj, 135, 512

\bibitem[{{Okabe} {et~al.}(2016){Okabe}, {Umetsu}, {Tamura}, {Fujita},
  {Takizawa}, {Matsushita}, {Fukazawa}, {Futamase}, {Kawaharada}, {Miyazaki},
  {Mochizuki}, {Nakazawa}, {Ohashi}, {Ota}, {Sasaki}, {Sato}, \& {Tam}}]{OKA16}
{Okabe}, N., {Umetsu}, K., {Tamura}, T., {et~al.} 2016, \mnras, 456, 4475

\bibitem[{P{\^a}ris {et~al.}(2017)P{\^a}ris, Petitjean, Ross, Myers, Aubourg,
  Streblyanska, Bailey, Armengaud, Palanque-Delabrouille, Y{\`e}che,
  {et~al.}}]{PAR17}
P{\^a}ris, I., Petitjean, P., Ross, N.~P., {et~al.} 2017, Astronomy \&
  Astrophysics, 597, A79

\bibitem[{{Pascale} {et~al.}(2011){Pascale}, {Auld}, {Dariush}, {Dunne},
  {Eales}, {Maddox}, {Panuzzo}, {Pohlen}, {Smith}, {Buttiglione}, {Cava},
  {Clements}, {Cooray}, {Dye}, {de Zotti}, {Fritz}, {Hopwood}, {Ibar},
  {Ivison}, {Jarvis}, {Leeuw}, {L{\'o}pez-Caniego}, {Rigby}, {Rodighiero},
  {Scott}, {Smith}, {Temi}, {Vaccari}, \& {Valtchanov}}]{PAS11}
{Pascale}, E., {Auld}, R., {Dariush}, A., {et~al.} 2011, \mnras, 415, 911

\bibitem[{P\'erez \& Granger(2007)}]{ipython}
P\'erez, F. \& Granger, B.~E. 2007, Computing in Science and Engineering, 9, 21

\bibitem[{{Pilbratt} {et~al.}(2010){Pilbratt}, {Riedinger}, {Passvogel},
  {Crone}, {Doyle}, {Gageur}, {Heras}, {Jewell}, {Metcalfe}, {Ott}, \&
  {Schmidt}}]{PIL10}
{Pilbratt}, G.~L., {Riedinger}, J.~R., {Passvogel}, T., {et~al.} 2010, \aap,
  518, L1

\bibitem[{{Planck Collaboration} {et~al.}(2014){Planck Collaboration}, {Ade},
  {Aghanim}, {Armitage-Caplan}, {Arnaud}, {Ashdown}, {Atrio-Barandela},
  {Aumont}, {Baccigalupi}, {Banday}, \& et~al.}]{Pla14}
{Planck Collaboration}, {Ade}, P.~A.~R., {Aghanim}, N., {et~al.} 2014, \aap,
  571, A19

\bibitem[{{Planck Collaboration} {et~al.}(2016){Planck Collaboration}, {Ade},
  {Aghanim}, {Arnaud}, {Ashdown}, {Aumont}, {Baccigalupi}, {Banday},
  {Barreiro}, {Bartolo}, \& et~al.}]{Pla16b}
{Planck Collaboration}, {Ade}, P.~A.~R., {Aghanim}, N., {et~al.} 2016, \aap,
  594, A21

\bibitem[{{Planck Collaboration} {et~al.}(2021){Planck Collaboration},
  {Aghanim}, {Akrami}, {Ashdown}, {Aumont}, {Baccigalupi}, {Ballardini},
  {Banday}, {Barreiro}, {Bartolo}, {Basak}, {Battye}, {Benabed}, {Bernard},
  {Bersanelli}, {Bielewicz}, {Bock}, {Bond}, {Borrill}, {Bouchet}, {Boulanger},
  {Bucher}, {Burigana}, {Butler}, {Calabrese}, {Cardoso}, {Carron},
  {Challinor}, {Chiang}, {Chluba}, {Colombo}, {Combet}, {Contreras}, {Crill},
  {Cuttaia}, {de Bernardis}, {de Zotti}, {Delabrouille}, {Delouis}, {Di
  Valentino}, {Diego}, {Dor{\'e}}, {Douspis}, {Ducout}, {Dupac}, {Dusini},
  {Efstathiou}, {Elsner}, {En{\ss}lin}, {Eriksen}, {Fantaye}, {Farhang},
  {Fergusson}, {Fernandez-Cobos}, {Finelli}, {Forastieri}, {Frailis},
  {Fraisse}, {Franceschi}, {Frolov}, {Galeotta}, {Galli}, {Ganga},
  {G{\'e}nova-Santos}, {Gerbino}, {Ghosh}, {Gonz{\'a}lez-Nuevo}, {G{\'o}rski},
  {Gratton}, {Gruppuso}, {Gudmundsson}, {Hamann}, {Handley}, {Hansen},
  {Herranz}, {Hildebrandt}, {Hivon}, {Huang}, {Jaffe}, {Jones}, {Karakci},
  {Keih{\"a}nen}, {Keskitalo}, {Kiiveri}, {Kim}, {Kisner}, {Knox},
  {Krachmalnicoff}, {Kunz}, {Kurki-Suonio}, {Lagache}, {Lamarre}, {Lasenby},
  {Lattanzi}, {Lawrence}, {Le Jeune}, {Lemos}, {Lesgourgues}, {Levrier},
  {Lewis}, {Liguori}, {Lilje}, {Lilley}, {Lindholm}, {L{\'o}pez-Caniego},
  {Lubin}, {Ma}, {Mac{\'\i}as-P{\'e}rez}, {Maggio}, {Maino}, {Mandolesi},
  {Mangilli}, {Marcos-Caballero}, {Maris}, {Martin}, {Martinelli},
  {Mart{\'\i}nez-Gonz{\'a}lez}, {Matarrese}, {Mauri}, {McEwen}, {Meinhold},
  {Melchiorri}, {Mennella}, {Migliaccio}, {Millea}, {Mitra},
  {Miville-Desch{\^e}nes}, {Molinari}, {Montier}, {Morgante}, {Moss}, {Natoli},
  {N{\o}rgaard-Nielsen}, {Pagano}, {Paoletti}, {Partridge}, {Patanchon},
  {Peiris}, {Perrotta}, {Pettorino}, {Piacentini}, {Polastri}, {Polenta},
  {Puget}, {Rachen}, {Reinecke}, {Remazeilles}, {Renzi}, {Rocha}, {Rosset},
  {Roudier}, {Rubi{\~n}o-Mart{\'\i}n}, {Ruiz-Granados}, {Salvati}, {Sandri},
  {Savelainen}, {Scott}, {Shellard}, {Sirignano}, {Sirri}, {Spencer},
  {Sunyaev}, {Suur-Uski}, {Tauber}, {Tavagnacco}, {Tenti}, {Toffolatti},
  {Tomasi}, {Trombetti}, {Valenziano}, {Valiviita}, {Van Tent}, {Vibert},
  {Vielva}, {Villa}, {Vittorio}, {Wandelt}, {Wehus}, {White}, {White},
  {Zacchei}, \& {Zonca}}]{PLA18_VI}
{Planck Collaboration}, {Aghanim}, N., {Akrami}, Y., {et~al.} 2021, \aap, 652,
  C4

\bibitem[{{Poglitsch} {et~al.}(2010){Poglitsch}, {Waelkens}, {Geis},
  {Feuchtgruber}, {Vandenbussche}, {Rodriguez}, {Krause}, {Renotte}, {van
  Hoof}, {Saraceno}, {Cepa}, {Kerschbaum}, {Agn{\`e}se}, {Ali}, {Altieri},
  {Andreani}, {Augueres}, {Balog}, {Barl}, {Bauer}, {Belbachir}, {Benedettini},
  {Billot}, {Boulade}, {Bischof}, {Blommaert}, {Callut}, {Cara}, {Cerulli},
  {Cesarsky}, {Contursi}, {Creten}, {De Meester}, {Doublier}, {Doumayrou},
  {Duband }, {Exter}, {Genzel}, {Gillis}, {Gr{\"o}zinger}, {Henning},
  {Herreros}, {Huygen}, {Inguscio}, {Jakob}, {Jamar}, {Jean}, {de Jong},
  {Katterloher}, {Kiss}, {Klaas}, {Lemke}, {Lutz}, {Madden}, {Marquet},
  {Martignac}, {Mazy}, {Merken}, {Montfort}, {Morbidelli}, {M{\"u}ller},
  {Nielbock}, {Okumura}, {Orfei}, {Ottensamer}, {Pezzuto}, {Popesso},
  {Putzeys}, {Regibo}, {Reveret}, {Royer}, {Sauvage}, {Schreiber}, {Stegmaier},
  {Schmitt}, {Schubert}, {Sturm}, {Thiel}, {Tofani}, {Vavrek}, {Wetzstein},
  {Wieprecht}, \& {Wiezorrek}}]{POG10}
{Poglitsch}, A., {Waelkens}, C., {Geis}, N., {et~al.} 2010, \aap, 518, L2

\bibitem[{{Rigby} {et~al.}(2011){Rigby}, {Maddox}, {Dunne}, {Negrello},
  {Smith}, {Gonz{\'a}lez-Nuevo}, {Herranz}, {L{\'o}pez-Caniego}, {Auld},
  {Buttiglione}, {Baes}, {Cava}, {Cooray}, {Clements}, {Dariush}, {de Zotti},
  {Dye}, {Eales}, {Frayer}, {Fritz}, {Hopwood}, {Ibar}, {Ivison}, {Jarvis},
  {Panuzzo}, {Pascale}, {Pohlen}, {Rodighiero}, {Serjeant}, {Temi}, \&
  {Thompson}}]{RIG11}
{Rigby}, E.~E., {Maddox}, S.~J., {Dunne}, L., {et~al.} 2011, \mnras, 415, 2336

\bibitem[{Ross {et~al.}(2012)Ross, Myers, Sheldon, Y{\`e}che, Strauss, Bovy,
  Kirkpatrick, Richards, Aubourg, Blanton, {et~al.}}]{ROS12}
Ross, N.~P., Myers, A.~D., Sheldon, E.~S., {et~al.} 2012, The Astrophysical
  Journal Supplement Series, 199, 3

\bibitem[{Schneider {et~al.}(2010)Schneider, Richards, Hall, Strauss, Anderson,
  Boroson, Ross, Shen, Brandt, Fan, {et~al.}}]{SCH10}
Schneider, D.~P., Richards, G.~T., Hall, P.~B., {et~al.} 2010, The Astronomical
  Journal, 139, 2360

\bibitem[{Schneider {et~al.}(2006)Schneider, Kochanek, \& Wambsganss}]{SCH06}
Schneider, P., Kochanek, C., \& Wambsganss, J. 2006, Gravitational lensing:
  strong, weak and micro: Saas-Fee advanced course 33, Vol.~33 (Springer
  Science \& Business Media)

\bibitem[{{Scranton} {et~al.}(2005){Scranton}, {M{\'e}nard}, {Richards},
  {Nichol}, {Myers}, {Jain}, {Gray}, {Bartelmann}, {Brunner}, {Connolly},
  {Gunn}, {Sheth}, {Bahcall}, {Brinkman}, {Loveday}, {Schneider}, {Thakar}, \&
  {York}}]{Scr05}
{Scranton}, R., {M{\'e}nard}, B., {Richards}, G.~T., {et~al.} 2005, \apj, 633,
  589

\bibitem[{{Smith} {et~al.}(2017){Smith}, {Ibar}, {Maddox}, {Valiante}, {Dunne},
  {Eales}, {Dye}, {Furlanetto}, {Bourne}, {Cigan}, {Ivison}, {Gomez}, {Smith},
  \& {Viaene}}]{SMI17}
{Smith}, M. W.~L., {Ibar}, E., {Maddox}, S.~J., {et~al.} 2017, \apjs, 233, 26

\bibitem[{{Stil} {et~al.}(2014){Stil}, {Keller}, {George}, \&
  {Taylor}}]{Stil14}
{Stil}, J.~M., {Keller}, B.~W., {George}, S.~J., \& {Taylor}, A.~R. 2014, \apj,
  787, 99

\bibitem[{{Swinbank} {et~al.}(2010){Swinbank}, {Smail}, {Longmore}, {Harris},
  {Baker}, {De Breuck}, {Richard}, {Edge}, {Ivison}, {Blundell}, {Coppin},
  {Cox}, {Gurwell}, {Hainline}, {Krips}, {Lundgren}, {Neri}, {Siana},
  {Siringo}, {Stark}, {Wilner}, \& {Younger}}]{SWI10}
{Swinbank}, A.~M., {Smail}, I., {Longmore}, S., {et~al.} 2010, \nat, 464, 733

\bibitem[{{Valiante} {et~al.}(2016){Valiante}, {Smith}, {Eales}, {Maddox},
  {Ibar}, {Hopwood}, {Dunne}, {Cigan}, {Dye}, {Pascale}, {Rigby}, {Bourne},
  {Furlanetto}, \& {Ivison}}]{VAL16}
{Valiante}, E., {Smith}, M.~W.~L., {Eales}, S., {et~al.} 2016, \mnras, 462,
  3146

\bibitem[{{Wang} {et~al.}(2011){Wang}, {Cooray}, {Farrah}, {Amblard}, {Auld},
  {Bock}, {Brisbin}, {Burgarella}, {Chanial}, {Clements}, {Eales},
  {Franceschini}, {Glenn}, {Gong}, {Griffin}, {Heinis}, {Ibar}, {Ivison},
  {Mortier}, {Oliver}, {Page}, {Papageorgiou}, {Pearson}, {P{\'e}rez-Fournon},
  {Pohlen}, {Rawlings}, {Raymond}, {Rodighiero}, {Roseboom}, {Rowan-Robinson},
  {Scott}, {Serra}, {Seymour}, {Smith}, {Symeonidis}, {Tugwell}, {Vaccari},
  {Vieira}, {Vigroux}, \& {Wright}}]{WAN11}
{Wang}, L., {Cooray}, A., {Farrah}, D., {et~al.} 2011, \mnras, 414, 596

\bibitem[{{Welikala} {et~al.}(2016){Welikala}, {B{\'e}thermin}, {Guery},
  {Strandet}, {Aird}, {Aravena}, {Ashby}, {Bothwell}, {Beelen}, {Bleem}, {de
  Breuck}, {Brodwin}, {Carlstrom}, {Chapman}, {Crawford}, {Dole}, {Dor{\'e}},
  {Everett}, {Flores-Cacho}, {Gonzalez}, {Gonz{\'a}lez-Nuevo}, {Greve},
  {Gullberg}, {Hezaveh}, {Holder}, {Holzapfel}, {Keisler}, {Lagache}, {Ma},
  {Malkan}, {Marrone}, {Mocanu}, {Montier}, {Murphy}, {Nesvadba}, {Omont},
  {Pointecouteau}, {Puget}, {Reichardt}, {Rotermund}, {Scott}, {Serra},
  {Spilker}, {Stalder}, {Stark}, {Story}, {Vanderlinde}, {Vieira}, \&
  {Wei{\ss}}}]{Wel16}
{Welikala}, N., {B{\'e}thermin}, M., {Guery}, D., {et~al.} 2016, \mnras, 455,
  1629

\bibitem[{{York} {et~al.}(2000){York}, {Adelman}, {Anderson}, {Anderson},
  {Annis}, {Bahcall}, {Bakken}, {Barkhouser}, {Bastian}, {Berman}, {Boroski},
  {Bracker}, {Briegel}, {Briggs}, {Brinkmann}, {Brunner}, {Burles}, {Carey},
  {Carr}, {Castander}, {Chen}, {Colestock}, {Connolly}, {Crocker}, {Csabai},
  {Czarapata}, {Davis}, {Doi}, {Dombeck}, {Eisenstein}, {Ellman}, {Elms},
  {Evans}, {Fan}, {Federwitz}, {Fiscelli}, {Friedman}, {Frieman}, {Fukugita},
  {Gillespie}, {Gunn}, {Gurbani}, {de Haas}, {Haldeman}, {Harris}, {Hayes},
  {Heckman}, {Hennessy}, {Hindsley}, {Holm}, {Holmgren}, {Huang}, {Hull},
  {Husby}, {Ichikawa}, {Ichikawa}, {Ivezi{\'c}}, {Kent}, {Kim}, {Kinney},
  {Klaene}, {Kleinman}, {Kleinman}, {Knapp}, {Korienek}, {Kron}, {Kunszt},
  {Lamb}, {Lee}, {Leger}, {Limmongkol}, {Lindenmeyer}, {Long}, {Loomis},
  {Loveday}, {Lucinio}, {Lupton}, {MacKinnon}, {Mannery}, {Mantsch}, {Margon},
  {McGehee}, {McKay}, {Meiksin}, {Merelli}, {Monet}, {Munn}, {Narayanan},
  {Nash}, {Neilsen}, {Neswold}, {Newberg}, {Nichol}, {Nicinski}, {Nonino},
  {Okada}, {Okamura}, {Ostriker}, {Owen}, {Pauls}, {Peoples}, {Peterson},
  {Petravick}, {Pier}, {Pope}, {Pordes}, {Prosapio}, {Rechenmacher}, {Quinn},
  {Richards}, {Richmond}, {Rivetta}, {Rockosi}, {Ruthmansdorfer}, {Sandford},
  {Schlegel}, {Schneider}, {Sekiguchi}, {Sergey}, {Shimasaku}, {Siegmund},
  {Smee}, {Smith}, {Snedden}, {Stone}, {Stoughton}, {Strauss}, {Stubbs},
  {SubbaRao}, {Szalay}, {Szapudi}, {Szokoly}, {Thakar}, {Tremonti}, {Tucker},
  {Uomoto}, {Vanden Berk}, {Vogeley}, {Waddell}, {Wang}, {Watanabe},
  {Weinberg}, {Yanny}, {Yasuda}, \& {SDSS Collaboration}}]{YOR00Y}
{York}, D.~G., {Adelman}, J., {Anderson}, John~E., J., {et~al.} 2000, \aj, 120,
  1579

\end{thebibliography}

\end{document}